\documentclass[%
reprint,
 amsmath,amssymb,
rmp,
]{revtex4-1}

\usepackage{graphicx}
\usepackage{dcolumn}
\usepackage{bm}
\usepackage{hyperref}

\begin{document}

\preprint{}

\title{Tutorials on X-ray Phase Contrast Imaging: Some Fundamentals and Some Conjectures on Future Developments}

\author{David M. Paganin}
 \affiliation{School of Physics and Astronomy, Monash University, Victoria 3800, Australia}
 \email{david.paganin@monash.edu}
\author{Daniele Pelliccia}%
\affiliation{Instruments \& Data Tools Pty Ltd, Victoria 3178, Australia}%
\email{daniele@idtools.com.au}


\begin{abstract}
These tutorials introduce some basics of imaging with coherent X-rays, focusing on phase contrast. We consider the transport-of-intensity equation, as one particular method for X-ray phase contrast imaging among many, before passing on to the inverse problem of phase retrieval.  These ideas are applied to two-dimensional and three-dimensional propagation-based phase-contrast imaging using coherent X-rays. We then consider the role of partial coherence, and sketch a generic means by which partially coherent X-ray imaging scenarios may be modelled, using the space--frequency description of partial coherence. Besides covering fundamental concepts in both theory and practice, we also give opinions on future trends in X-ray phase contrast imaging including X-ray tomography, and comparison of different phase contrast imaging methods. These tutorials will be accessible to those with a basic background in optics ({\em e.g.}~wave equation, Maxwell equations, Fresnel and Fraunhofer diffraction, and the basics of Fourier and vector analysis) and interactions of X-rays with matter ({\em e.g.}~attenuation mechanisms and complex refractive index).
 \newline\newline Fifteen video lectures, based directly on these notes, are at: \url{https://bit.ly/2GdoVg8}
 \newline\newline {\em We humbly dedicate these notes to the memory of Claudio Ferrero.} 
 \end{abstract}

\maketitle

\tableofcontents

\section{Introduction to these tutorials}

We consider the field of X-ray phase contrast imaging from a tutorial perspective.  We cover some basics of the field, augmenting our discussions throughout with speculations regarding future lines of development of the field.  The primary intended audience is those commencing research in the field, although it is our intention that these largely self-contained notes be more broadly accessible. 

Each of the three parts of our tutorial commences with a theory component focused upon the mathematical-physics underpinning of the topics treated therein.  The first two parts are rounded out with a complementary component giving practical applications and examples. 

The first part of these notes deals with X-ray imaging basics, sketching a passage from the Maxwell equations of classical electrodynamics, through to the paraxial wave equation describing coherent scalar X-ray fields.  We also introduce the projection approximation, Fresnel diffraction, absorption contrast and phase contrast.  We then examine, from a practical perspective, the validity conditions of the projection approximation, including the conditions under which this approximation is likely to break down.  Some attention is also given to the question of tomography beyond the projection approximation, including the roles of diffraction tomography and the multi-slice approach.

The second part deals with elements of X-ray phase contrast imaging and the associated inverse problem of phase retrieval. We begin with an outline of the transport-of-intensity equation, which is tied to one of the common phase-contrast methods, namely propagation-based X-ray phase contrast.  Rather than subsequently considering in detail a multiplicity of other powerful methods for X-ray phase contrast imaging, we instead generalise a wide class of such phase contrast imaging systems, by considering many of them to be particular examples of shift-invariant coherent linear imaging systems. We give some time to considering the associated transfer function concept, and the realisation of X-ray phase contrast imaging in such a general setting.  We indicate some key concepts in the underpinning theory of forward problems and inverse problems, before considering the inverse problem of phase retrieval.  Two particular examples of phase retrieval are briefly considered, namely transport-of-intensity phase retrieval in both two and three spatial dimensions.  The practice component then gives broader consideration to the suite of available X-ray phase contrast imaging methods, and discusses some similarities between certain of these methods.  We emphasise that no one method of X-ray phase contrast imaging is superior to all others in all circumstances, arguing rather that each have their relative strengths and weaknesses.

The third and final part considers partial coherence, with particular reference to X-ray phase contrast imaging using partially coherent radiation processed via arbitrary linear imaging systems.  We seek to give a general means to theoretically and computationally model a very large class of such X-ray phase contrast imaging systems, both those that currently exist, and many of those that may be developed in the future.  The key underpinning idea is the space--frequency description of partial coherence, whereby one has a statistical ensemble of strictly monochromatic fields at each spatial frequency, independently propagating through a given generalised X-ray phase contrast imaging system.  This allows one to determine, in an efficient manner, the resulting spectral density ({\em i.e.}~ensemble averaged intensity) at any point in the imaging system, together with the transport through the system of coherence functions such as the cross-spectral density, the Wigner function and the ambiguity function. Lastly, we give a number of opinions and speculations regarding possible future developments in the field.  

Further detail, on many of the topics presented here, is available in the text by Paganin (2006).  Note also that these tutorials do not claim in any way to be a representative overview of the field.  Rather, they are intended as an introductory overview of certain key aspects of the field, which contains enough entry points to the published literature to empower a journey of further exploration.  

\part{X-ray imaging basics}

\section{Theory}

We cover some basics of coherent X-ray imaging, including the wave equations for X-ray waves and their interactions with matter, the projection approximation, Fresnel diffraction and phase contrast.  

\subsection{Vector vacuum wave equations} 

The Maxwell equations, which govern the evolution of classical electromagnetic fields in space and time, lead to the following d'Alembert equations for the electric field ${\bf E}(x,y,z,t)$ and magnetic field ${\bf B}(x,y,z,t)$ in free space:
\begin{eqnarray}\label{eq:WaveEquationsElectric}
\left(\frac{1}{c^2}\frac{\partial^2}{\partial
t^2}-\nabla^2\right){\bf E}(x,y,z,t)={\bf 0},
\\ \label{eq:WaveEquationsMagnetic}\left(\frac{1}{c^2}\frac{\partial^2}{\partial
t^2}-\nabla^2\right){\bf B}(x,y,z,t)={\bf 0}.
\end{eqnarray}
\noindent Here, $(x,y,z)$ are Cartesian spatial coordinates, $t$ is time, $c$ is a speed given by
\begin{equation}\label{eq:SpeedOfLight}
    c=\frac{1}{\sqrt{\mu_0\varepsilon_0}},
\end{equation}
\noindent the Laplacian in three spatial dimensions is
\begin{equation}
\nabla^2=\frac{\partial^2}{\partial x^2}+\frac{\partial^2}{\partial y^2}+\frac{\partial^2}{\partial z^2}, 
\end{equation}
$\epsilon_0$ is the electrical permittivity of free space and $\mu_0$ is the magnetic permeability of free space.  Syst\`{e}me-Internationale (SI) units are used consistently. The notation used throughout is the same as Paganin (2006). 

The above equations imply two facts which were quite revolutionary when first discovered: (i) Electromagnetic disturbances propagate as waves in vacuum; (ii) the speed $c$ of these electromagnetic waves, given by the Maxwell relation (Eq.~\ref{eq:SpeedOfLight}), coincides so closely with the speed of light in vacuum, to suggest that {\em light is an electromagnetic wave.}  In the late nineteenth-century context in which it was derived, this then-radical observation unified what were previously thought to be three separate bodies of physics knowledge: electricity, magnetism and (visible light) optics.  

This was indeed a colossal moment in the history of physics. Before the advent of the Maxwell equations and the associated discovery that visible light is an electromagnetic disturbance, there were five separate theories describing aspects of the physical world: electricity, magnetism, (visible light) optics, thermodynamics and mechanics.  With both the Maxwell equations and the discovery that light is an electromagnetic wave, the first three of these theories united into one overarching theory of electromagnetism and electromagnetic waves.  Such a unification remains a guiding light in modern quests for a unification of quantum theory with Einstein's gravitational theory of general relativity. 

Returning to the main thread of our argument, we now know that the class of electromagnetic waves is not exhausted by those that are visible to the human eye.  Of particular focus to us are X-ray electromagnetic waves.

\subsection{Scalar vacuum wave equation \& complex wave-function} 

Equations~\ref{eq:WaveEquationsElectric} and \ref{eq:WaveEquationsMagnetic} are a pair of vector equations, or, equivalently, a set of six scalar equations: three for the Cartesian components of the electric field, and three for the Cartesian components of the magnetic field.  Each of these six scalar vacuum field equations has the form:  
\begin{eqnarray}\label{eq:WaveEquationComplexDisturbance}
\left(\frac{1}{c^2}\frac{\partial^2}{\partial
t^2}-\nabla^2\right)\Psi(x,y,z,t)=0.
\end{eqnarray}

It is convenient to treat $\Psi(x,y,z,t)$ as a complex function, termed the ``wave function'', which describes the X-ray field.  Only the real part of this wave-function is physically meaningful, but we will not need to make use of this fact at any point in these notes. By transitioning from a vector-wave description to a scalar-wave description of the X-ray field, polarisation is implicitly neglected (or a single linear polarisation is implicitly assumed). This assumption is often reasonable in many paraxial imaging and diffraction contexts.  However, note that there are many cases ({\em e.g.}~magnetic scattering of circularly-polarised X-rays, and dynamical diffraction from near-perfect crystals) where the effects of X-ray polarisation must be taken into account.

\subsection{Physical meaning of intensity and phase}

At each point $(x,y,z)$ in space, for each instant of time $t$, $\Psi(x,y,z,t)=0$ is a complex number.  Complex numbers have magnitude and phase, so we may write:
\begin{equation}\label{eq:MadelungDecomposition}
\Psi(x,y,z,t)=\sqrt{I(x,y,z,t)}\exp[i\phi(x,y,z,t)].
\end{equation}

The magnitude of $\Psi(x,y,z,t)$ has been written as $\sqrt{I(x,y,z,t)}$, so that
\begin{equation}
I(x,y,z,t)=|\Psi(x,y,z,t)|^2,
\end{equation}
\noindent where $I(x,y,z,t)$ is the intensity of the field.  The phase of $\Psi(x,y,z,t)=0$ has been denoted $\phi(x,y,z,t)$. For the instant of time $t$, surfaces of constant phase may be identified with wave-fronts of the X-ray field.  These fronts move at (extremely close to) the speed of light in vacuum\footnote{We are here alluding to the subtlety that the speed of light in vacuum is in general less than the speed of a plane wave in vacuum.  See Giovannini {\em et al.}~(2015) for further information on this fascinating point.}, in a direction that is typically away from the source generating those waves -- see Fig.~1(a).   

\begin{figure}
\includegraphics[scale=0.30]{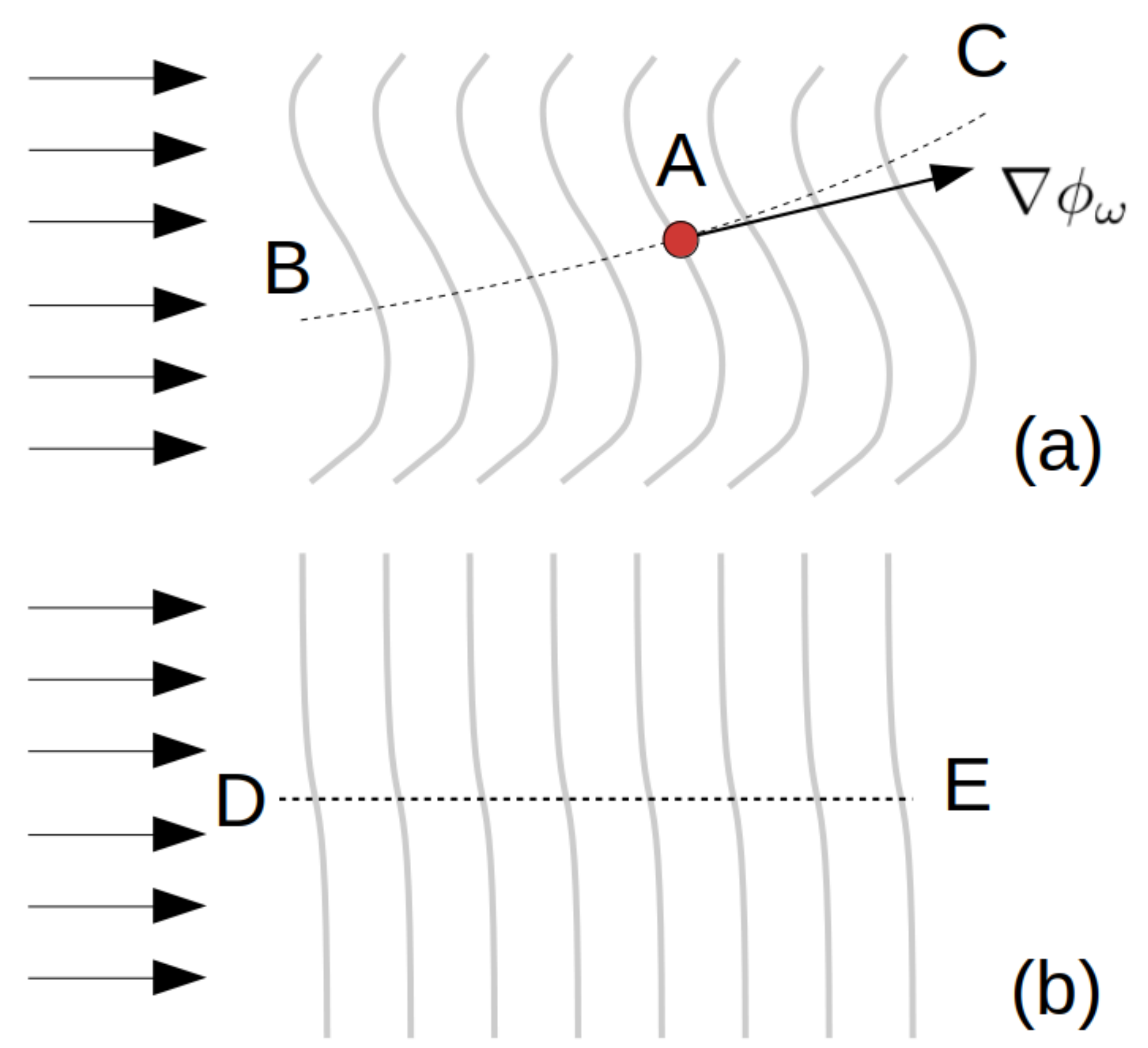}
\caption{X-ray wave-fronts.  (a) At a given instant of time $t=t_0$, the intensity of an X-ray field is $I(x,y,z,t=t_0)$.  The wave-fronts, namely the surfaces of constant phase $\phi(x,y,z,t=t_0)$, are indicated by the series of curved surfaces.  At any point $A$ the wave-fronts move away from the source as time increases.  The direction of energy flow at point $A$ is $\nabla\phi(x,y,z,t=t_0)$.  The associated current density (Poynting vector) is ${\bf S}(x,y,z,t=t_0)\propto I(x,y,z,t=t_0) \nabla \phi(x,y,z,t=t_0)$, which is perpendicular to each wave-front.  The current density is everywhere tangent to the associated streamlines, such as the streamline $BAC$ passing through the point $A$.  (b) For a paraxial field, all wave-fronts may be obtained by slightly deforming planar surfaces perpendicular to the optic axis ({\em i.e.}~the $z$ axis, which runs from left to right).  All Poynting vectors are close to being {\bf par}allel to the optic {\bf ax}is (hence the term {\bf parax}ial). The associated streamlines such as $DE$ are well approximated by straight lines parallel to the optic axis.  Image adapted from Paganin (2006).}
\label{fig:XRayWaveFronts}
\end{figure}

\subsection{Fully coherent fields}

Assume the field to be strictly monochromatic, and therefore perfectly coherent, so that its time development at any point in space oscillates with a fixed angular frequency $\omega$:
\begin{equation}\label{eq:MonochromaticField}
\Psi(x,y,z,t)=\psi_{\omega}(x,y,z)\exp(-i\omega t),
\end{equation}
\noindent where
\begin{equation}
\omega = 2\pi f = ck,  
\end{equation}
\noindent $f$ denotes frequency, and $k$ is the wave-number corresponding to vacuum wavelength $\lambda$:
\begin{equation}
k=2\pi/\lambda.  
\end{equation}

Substitution of Eq.~\ref{eq:MonochromaticField} into Eq.~\ref{eq:WaveEquationComplexDisturbance} gives the {\em Helmholtz equation} in vacuum:
\begin{equation}\label{eq:Helmholtz_Equation}
    \left(\nabla^2+k^2\right)\psi_{\omega}(x,y,z)=0.
\end{equation}

This vacuum wave equation for coherent scalar electromagnetic waves may be generalised to account for the presence of material media.  Such media will be assumed to be static, non-magnetic, and sufficiently slowly spatially varying, so that they may be described by a position-dependent refractive index $n_{\omega}(x,y,z)$.  This refractive index alters the vacuum wavelength as follows:
\begin{equation}
\lambda\longrightarrow\frac{\lambda}{n_{\omega}(x,y,z)},
\end{equation}
\noindent hence 
\begin{equation}
k\longrightarrow k \, n_{\omega}(x,y,z). 
\end{equation}
The vacuum Helmholtz equation (Eq.~\ref{eq:Helmholtz_Equation}) therefore becomes the Helmholtz equation in the presence of non-magnetic scattering media:
\begin{eqnarray}\label{eq:ScalarHelmholtzEquationInRefractiveMedium}
    \left[\nabla^2+k^2n_{\omega}^2(x,y,z) \right]
    \psi_{\omega}(x,y,z)=0.
\end{eqnarray}
See {\em e.g.}~Paganin (2006) for a full derivation of the above equation, which elaborates on the key assumptions that the scattering medium be (i) linear, (ii) isotropic, (iii) static, (iv) non-magnetic, (v) have zero charge density and (vi) zero current density, and (vii) be spatially slowly varying in its material properties.

As an interesting aside, note that the above equation is mathematically identical in form to the time-independent Schr\"{o}dinger equation for non-relativistic electrons in the presence of a scalar scattering potential (this latter equation assumes that the effects of electron spin can be ignored, and that the material with which the electron interacts is non-magnetic).  Hence, the research fields of coherent X-ray optics and transmission electron microscopy have much in common.  Indeed, we are of the view that fundamental progress in both fields would  advance more rapidly if more workers from each field were to familiarise themselves with work from both fields.  

As a second aside, we recall the statement invoked in deriving Eq.~\ref{eq:ScalarHelmholtzEquationInRefractiveMedium}, namely the three assumptions that the scattering ``media will be assumed to be static, non-magnetic, and sufficiently slowly spatially varying, so that they may be described by a position-dependent refractive index''.  The breakdown of any or all of these three key assumptions leads to interesting generalisations that will not be considered here.  For example, 

(i) the breakdown of the first assumption enters us into the very interesting realm of time-dependent samples, including those that experience radiation damage during the act of X-ray imaging; 

(ii) the breakdown of the second assumption is key to the study of magnetic materials using, for example, circularly polarised X-rays; 

(iii) the breakdown of the third assumption will become progressively more important as X-ray imaging is pushed more and more often to regions of high resolution, {\em e.g.}~on nm and smaller length scales.

\subsection{Coherent paraxial fields} 

With reference to Fig.~1(b), assume our monochromatic complex scalar X-ray wave-field to be paraxial, in the sense described in the caption to the said figure.  Under this approximation it is natural to express the complex disturbance ${\psi}_{\omega}(x,y,z)$ as a product of a $z$-directed plane wave $\exp(ikz)$, and a perturbing envelope  $\tilde{\psi}_{\omega}(x,y,z)$.  Without any loss of generality, we then have:
\begin{equation}\label{eq:EnvelopeForPlaneWave}
{\psi}_{\omega}(x,y,z)\equiv\tilde{\psi}_{\omega}(x,y,z)\exp(ikz).
\end{equation}

Conveniently, 
\begin{equation}
|\tilde{\psi}_{\omega}(x,y,z)|^2=|\psi_{\omega}(x,y,z)|^2=I_{\omega}(x,y,z), 
\end{equation}
\noindent so that the intensity of the envelope $\tilde{\psi}_{\omega}(x,y,z)$ is the same as the intensity of ${\psi}_{\omega}(x,y,z)$.

Now, if Eq.~\ref{eq:EnvelopeForPlaneWave} is substituted into Eq.~\ref{eq:ScalarHelmholtzEquationInRefractiveMedium}, and a term containing the second $z$ derivative is discarded as being small compared to the other terms on account of the paraxial assumption, one obtains: 
\begin{equation}\label{eq:ParaxialEquationInhomogeneous}
 \left(2ik\frac{\partial}{\partial z}+\nabla_{\perp}^2+k^2[n_{\omega}^2(x,y,z)-1]\right)\tilde{\psi}_{\omega}(x,y,z)=0,
\end{equation}
\noindent where
\begin{equation}
    \nabla_{\perp}^2\equiv\frac{\partial^2}{\partial x^2}+\frac{\partial^2}{\partial
    y^2}
\end{equation}
\noindent is the transverse Laplacian ({\em i.e.}, the Laplacian in the $(x,y)$ plane perpendicular to the optic axis $z$, so that $\nabla^2=\nabla_{\perp}^2+ \partial^2/ \partial z^2$).

We again draw a parallel with quantum mechanics, noting that Eq.~\ref{eq:ParaxialEquationInhomogeneous} is mathematically identical in form to the time-{\em dependent} Schr\"{o}dinger equation in 2+1 dimensions ({\em i.e.}~two space dimensions and one time dimension), in the presence of a time-dependent scalar potential $V(x,y,t)$, if one replaces $z$ with $t$, and considers $n_{\omega}^2(x,y,z\rightarrow t)-1$ to be proportional to $-V(x,y,t)$.   

\subsection{Projection approximation \& absorption contrast}

Consider Fig.~2.  Here, $z$-directed monochromatic complex scalar X-ray waves illuminate a static non-magnetic object, from the left.  By assumption, the object is totally contained within the slab of space between $z=0$ and $z=z_0\ge 0$.  The object is described by its refractive index distribution $n_{\omega}(x,y,z)$, which will only differ from unity ({\em i.e.}~the refractive index of vacuum) within the volume occupied by the object.

We wish to determine the complex disturbance (wave-function) over the plane $z=z_0$, which is termed the ``exit surface'' of the object, as a function of both (i) the complex disturbance over the ``entrance surface'' $z=0$ and (ii) the refractive index distribution of the object. 

\begin{figure}
\includegraphics[scale=0.2]{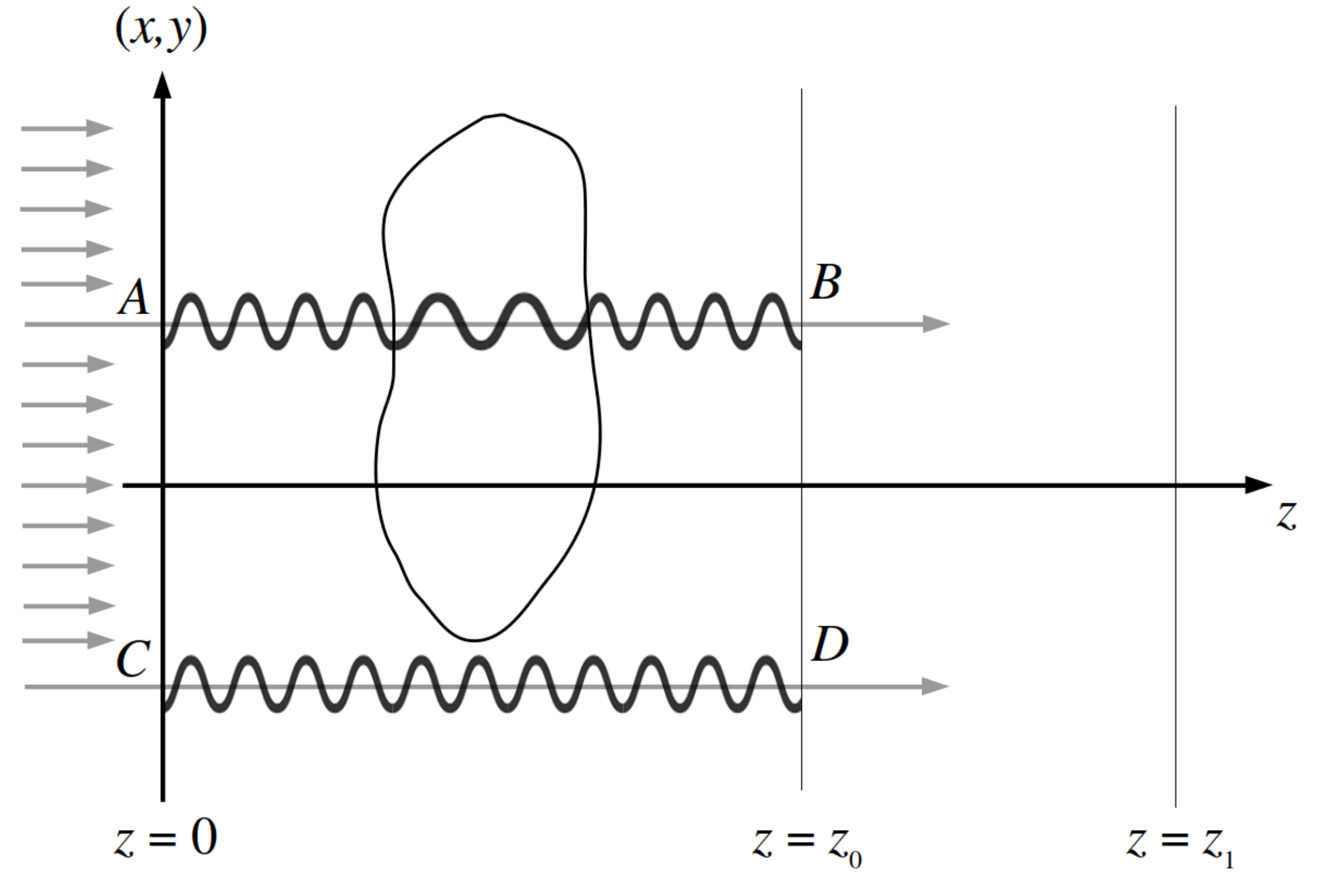}
\caption{From entrance-surface to exit-surface wave-field, under the projection approximation. Adapted from Paganin (2006).}
\end{figure}

We assume the object to be sufficiently slowly varying in space, that all streamlines of the X-ray flow may be well approximated by straight lines parallel to $z$.  Under this so-called {\em projection approximation}, the validity conditions for which are further discussed in Sec.~\ref{Sec:ValidityOfProjApprox} below, the transverse Laplacian may be neglected in Eq.~\ref{eq:ParaxialEquationInhomogeneous}.  Thus:
\begin{equation}
\frac{\partial}{\partial
z}\tilde{\psi}_{\omega}(x,y,z)\approx\frac{k}{2i}[1-n_{\omega}^2(x,y,z)]\tilde{\psi}_{\omega}(x,y,z).
\end{equation}

For each fixed point $(x,y)$, this is a simple linear first-order ordinary differential equation, which can be immediately integrated with respect to $z$ to give:
\begin{eqnarray}\label{eq:zz7}
    \tilde{\psi}_{\omega}(x,y,z=z_0) \approx \qquad\qquad\qquad\qquad\qquad\qquad\qquad\qquad \\ \nonumber \exp
    \left\{\frac{k}{2i}\int_{z=0}^{z=z_0}[1-n_{\omega}^2(x,y,z)]dz\right\}\tilde{\psi}_{\omega}(x,y,z=0).
\end{eqnarray}

At this point, it is convenient to introduce a complex form for the refractive index, the real part of which corresponds to the refractive index; we shall see that the imaginary part of this complexified refractive index can be related to the absorptive properties of a sample.  With this in mind, write the complex refractive index as
\begin{equation}\label{eq:ComplexRefractiveIndex}
    n_{\omega}=1-\delta_{\omega}+i\beta_{\omega},
\end{equation}

\noindent where 
\begin{equation}
|\delta_{\omega}|, |\beta_{\omega}| \ll 1
\end{equation}
since the complex refractive index for hard X-rays is typically extremely close to unity.  Hence:
\begin{eqnarray}\label{eq:OneMinusRefrIndexSquaredApproximate}
  1-n_{\omega}^2(x,y,z)\approx
  2[\delta_{\omega}(x,y,z)-i\beta_{\omega}(x,y,z)],
\end{eqnarray}
where we have discarded terms containing $\delta_{\omega}^2$, $\beta_{\omega}^2$ and $\delta_{\omega}\beta_{\omega}$ since these will be much smaller than the terms that have been retained in the right side of Eq.~\ref{eq:OneMinusRefrIndexSquaredApproximate}.

If the above expression is substituted into Eq.~\ref{eq:zz7}, we obtain %
\begin{eqnarray}\label{eq:zz8}
\tilde{\psi}_{\omega}(x,y,z=z_0)\approx  \tilde{\psi}_{\omega}(x,y,z=0) \qquad\qquad\qquad\qquad \\ \nonumber \times\exp
\left\{-ik\int_{z=0}^{z=z_0}[\delta_{\omega}(x,y,z)-i\beta_{\omega}(x,y,z)]dz\right\} \\ \nonumber.
\end{eqnarray}
This shows that the exit wave-field $\tilde{\psi}_{\omega}(x,y,z=z_0)$ may be obtained from the entrance wave-field $\tilde{\psi}_{\omega}(x,y,z=0)$ via multiplication by a {\em transmission function} ${\mathcal T}(x,y)$.  The transmission function is the second line of Eq.~\ref{eq:zz8}.

The position-dependent phase shift
\begin{equation}
\arg{\mathcal T}(x,y)\equiv \Delta\phi(x,y),
\end{equation}
due to the object, is:
\begin{equation}\label{eq:PhaseShiftProjectionApproximation}
    \Delta\phi(x,y)=-k\int\delta_{\omega}(x,y,z)dz.
\end{equation}

\noindent The above expression quantifies the deformation of the X-ray wave-fronts due to passage through the object.  Physically, for each fixed transverse coordinate $(x,y)$, phase shifts (and the associated wave-front deformations) are continuously accumulated along energy-flow streamlines (loosely, ``rays'') such as $AB$ in Fig.~2.  In making all of these statements, it is useful to look back to Fig.~1 and recall the direct connection between the phase of a complex wave-field, and its associated wave-fronts.  The phase shifts---associated with passage of an X-ray wave through an object---quantify the wave-front deformations and associated refractive properties of the object.  Also, since we are working with a wave picture rather than the less-general ray picture for X-ray light, refraction is associated with wave-front deformation rather than ray deflection.

Refraction, due to the object, is a property that may be augmented by the {\em attenuation} due to the object.  This latter quantity may be obtained by taking the squared modulus of Eq.~\ref{eq:zz8}, to give the Beer--Lambert law:
\begin{eqnarray}\label{eq:zz8a}
    I_{\omega}(x,y,z=z_0) \quad\quad\quad\quad\quad\quad\quad\quad\quad\quad\quad\quad \\ \nonumber=\exp\left[-\int\mu_{\omega}(x,y,z)dz\right]I_{\omega}(x,y,z=0). \end{eqnarray}
Above, we have used the following expression relating the imaginary part $\beta_{\omega}$ of the refractive index, to the associated {\em linear attenuation coefficient} $\mu_{\omega}$: 
\begin{equation}
\label{eq:AbsorptionCoefficient}
    \mu_{\omega}=2k\beta_{\omega}.
   \end{equation}

Note for later reference, that Eq.~\ref{eq:zz8a} may also be written in the logarithmic form:
\begin{eqnarray}\label{eq:zz8a_log_form}
    \log_e [I_{\omega}(x,y,z=z_0)/I_{\omega}(x,y,z=0)] \quad\quad\quad \\ \nonumber=-\int\mu_{\omega}(x,y,z)dz.
\end{eqnarray}

Equation~\ref{eq:zz8a} forms the basis for {\em absorption contrast imaging}.  In particular, if a two-dimensional position sensitive detector is placed in the plane $z=z_0$ in Fig.~2, and the illuminating radiation has an intensity $I_{\omega}(x,y,z=0)$ that is approximately constant with respect to $x$ and $y$, then all contrast in the resulting ``contact'' image will be due to local absorption of rays such as $AB$ in Fig.~2.  While the logarithm of this image is sensitive to the projected linear attenuation coefficient $\int\mu_{\omega}(x,y,z)dz$, the contact image contains no contrast whatsoever, due to the phase shifts quantified by Eq.~\ref{eq:PhaseShiftProjectionApproximation}.  This lack of {\em phase contrast}, in conventional contact X-ray imaging, is unfortunate, since many structures of interest (such as soft biological tissues) are close to being non-absorbing, meaning that they are poorly visualised, if at all, in absorption-contrast X-ray imaging.     

\subsection{Fresnel diffraction \& propagation-based phase contrast}

Consider Fig.~3, which shows a source $A$ radiating into space.  Optical elements and samples, which may lie between $A$ and the plane $z=0$, are not shown. The {\em diffraction problem} seeks to determine the wave-field over the plane $z>0$, given the disturbance over the plane $z=0$.  The space $z\ge 0$ is assumed to be vacuum, and all waves in this space are assumed to be both paraxial with respect to the optic axis $z$, and monochromatic.

\begin{figure}
\includegraphics[scale=0.3]{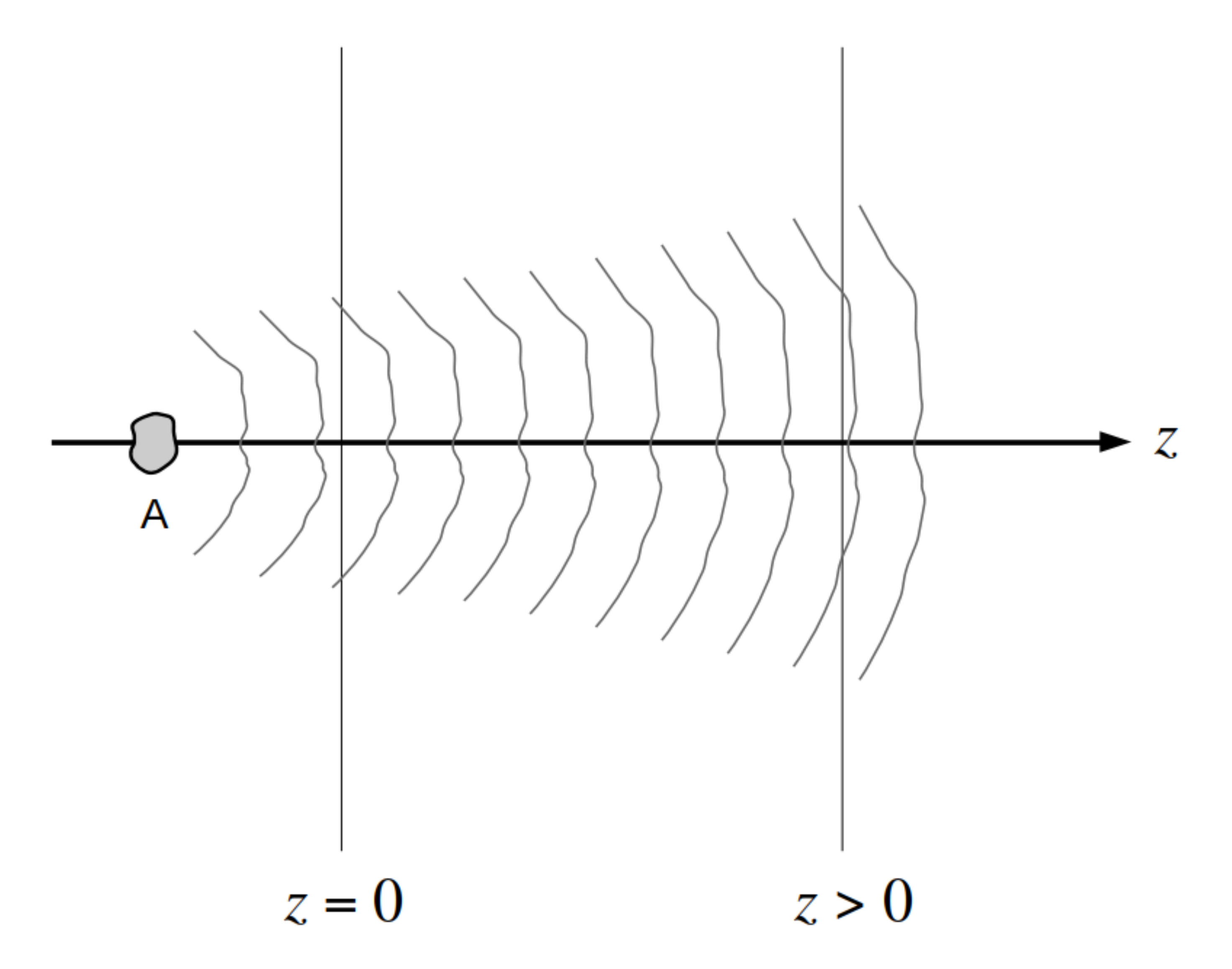}
\caption{Free-space propagation of X-ray waves. Image adapted from Paganin (2006).}
\end{figure}

In the space $z\ge0$ the waves will obey the ``$n_{\omega}(x,y,z)=1$'' special case of Eq.~\ref{eq:ParaxialEquationInhomogeneous}, namely:
\begin{equation}\label{eq:ParaxialEquation}
 \left(2ik\frac{\partial}{\partial
 z}+\nabla_{\perp}^2\right)\tilde{\psi}_{\omega}(x,y,z)=0.
\end{equation}

The solution to the diffraction problem, based on the above {\em free-space paraxial equation}, may be written as:
\begin{equation}\label{eq:FresnelDiffraction}
    \psi_{\omega}(x,y,z=\Delta)={\mathcal{D}}_{\Delta}\psi_{\omega}(x,y,z=0),\hspace{1em} \Delta\ge0.
\end{equation}
\noindent Here, ${\mathcal{D}}_{\Delta}$ is a (Fresnel) diffraction operator, which acts on the unpropagated forward-travelling field $\psi_{\omega}(x,y,z=0)$, propagating it a distance $\Delta$, to give $\psi_{\omega}(x,y,z=\Delta)$.  An expression for ${\mathcal{D}}_{\Delta}$, which may be readily derived from the free-space paraxial equation, will be given later.  

From the squared magnitude of Eq.~\ref{eq:FresnelDiffraction}, it is clear that the {\em intensity of the propagated field depends on both the intensity and phase of the unpropagated field.}  This point is both trivial---because the right side, of the squared modulus of Eq.~\ref{eq:FresnelDiffraction}, obviously depends on the phase---and of profound importance, since it implies that {\em the Fresnel diffraction pattern, namely the propagated intensity over the plane $z > 0$ in Fig.~3, provides the phase contrast that was missing from the contact image.}  

This mechanism, for obtaining intensity contrast (in the plane $z>0$) that is sensitive to phase variations (in the plane $z=0$), is known as {\em propagation based phase contrast}. This phenomenon has been known under different names for millennia in visible-light optics, {\em e.g.}~from the heat shimmer over a hot road, and known for many decades in both the visible-light microscopy ({\em e.g.}~Zernike, 1942; Bremmer, 1952) and electron microscopy ({\em e.g.}~Cowley, 1959) communities.  The X-ray community only became significantly aware of this phenomenon from the early 1990s ({\em e.g.}~White and Cerrina, 1992), including pioneering studies by a number of scientists from the European Synchrotron, such as those of Snigirev and colleagues (1995), and Cloetens and colleagues (1996).  Other noteworthy X-ray papers from this period include those of Wilkins and colleagues (1996), and Nugent and colleagues (1996).         

For the remainder of this subsection, we seek to further develop the intuition of the reader, regarding the qualitative nature of propagation-based X-ray phase contrast. 

\begin{figure}
\includegraphics[scale=0.2]{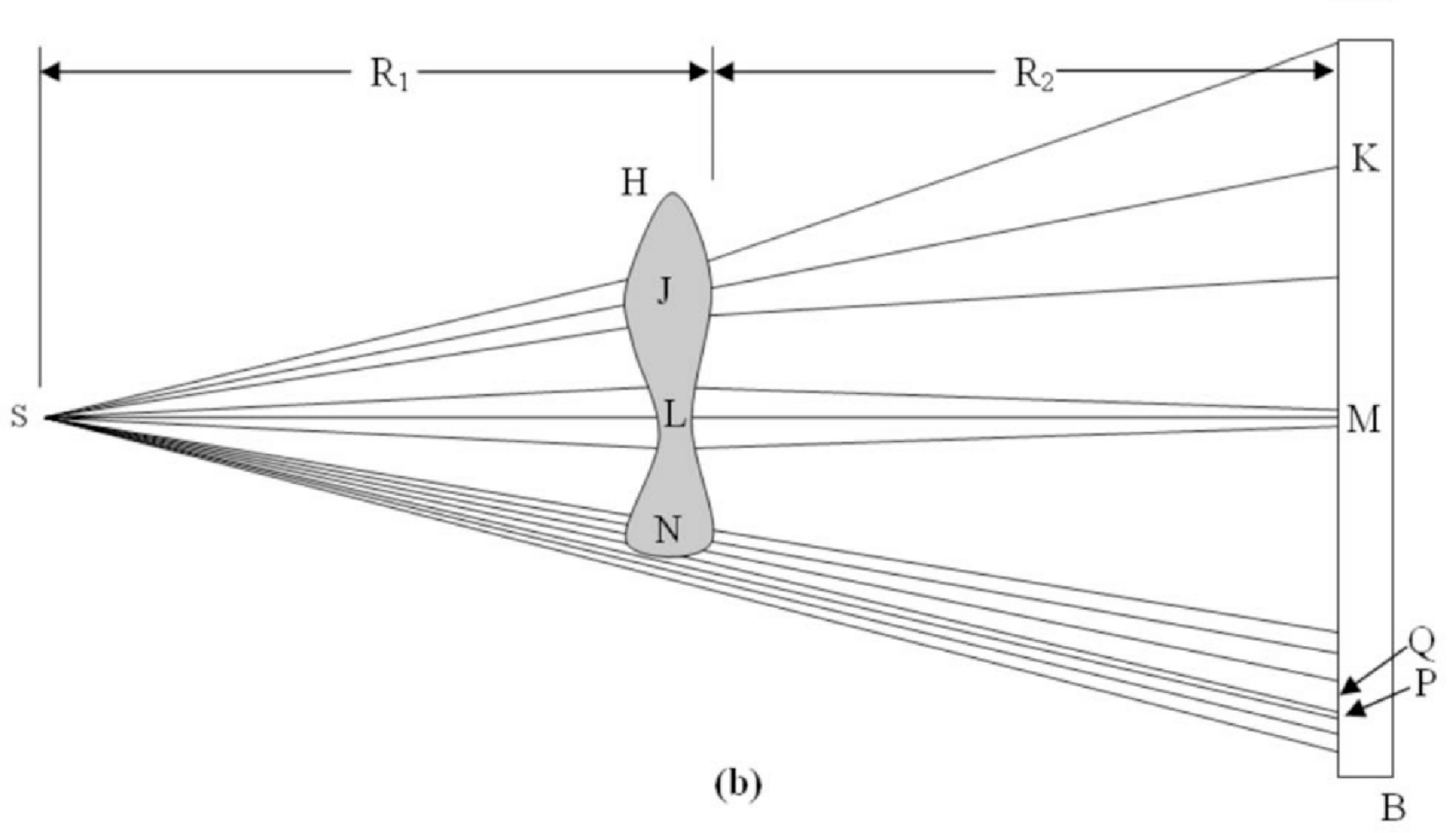}
\caption{Geometric-optics diagram to aid understanding of propagation-based X-ray phase contrast imaging.  Here, $S$ is a small X-ray source, which illuminates an object shown in grey.  The imaging plane is shown to the right of the image.  The source-to-object and object-to-detector distances are denoted by $R_1$ and $R_2$ respectively. Reproduced with permission from Gureyev {\em et al.}~(2009).}
\label{fig:RRR1}
\end{figure}

With this end in mind, consider Fig.~\ref{fig:RRR1}, in which a small X-ray source $S$ illuminates an object $H$ shown in grey.  The source-to-object distance is denoted by $R_1$ and the object-to-detector distance is denoted by $R_2$.  The distance $R_2$ is assumed to be large enough that propagation-based phase contrast is manifest over the detector plane $B$, but not so large that multiple Fresnel diffraction fringes are present\footnote{More precisely, we are here assuming the Fresnel number $N_F= M a^2/(\lambda R_2)$ to be much greater than unity. Here, $a$ corresponds to the smallest transverse characteristic feature size in the object that is not smeared out by the finite size of the source, $R_2$ is the object-to-detector distance and $M=(R_1+R_2)/R_1$ is the geometric magnification.  Note also that the concept of the Fresnel number is used in a different but related context later in these notes, when it is used to consider the conditions under which the projection approximation is valid for a given sample.}.  Assuming the object to be sufficiently thin that the projection approximation holds, one may identify three different features within the object, which are here labelled $J$, $L$ and $N$.  

\begin{itemize}

\item Features such as $J$ correspond to either the thin object in projection behaving locally like a convex lens, or a point within the volume of an object which has a local peak of density.  Because the real part of the complex refractive index is less than unity for X-rays, convex X-ray lenses are defocusing optical elements ({\em cf.}~the case for visible light, where convex lenses are focusing optics since the real part of the refractive index is greater than unity).  Since $J$ may be considered as a defocusing feature in the object, the local ray density at point $K$ on the detector will be lessened via the refractive effects of $J$; hence point $K$ in the detector will have reduced brightness, on account of the propagation distance $R_2$ that lies between $J$ and $K$.

\item Features such as $L$, which may be either a concave feature in the projected thickness of the sample or a feature within the sample which has a local trough of density, will act as a converging lens for X-rays.  Hence the intensity at point $M$ will be increased by the effects of refraction by feature $L$, provided that $R_2$ is large enough for the intensity-increasing effects of the focusing element $L$ to be manifest at point $M$ on the detector. Note the crucial role played by the object-to-detector distance $R_2$, through which the wave propagates before reaching the detector.

\item One also has propagation-based phase contrast due to features such as $N$, which correspond to points on the edge of the object.  Here, ``edge'' refers to the edge of the object when projected along the optic axis $z$.  On account of Fresnel diffraction in the slab of vacuum between the object and the detector, the propagation-based phase contrast signature of an edge such as $N$ will be an increase of intensity at point $P$, with a corresponding decrease at point $Q$.  Such ``edge contrast'' is a characteristic feature of propagation-based X-ray phase contrast.

\end{itemize}

\begin{figure}
\includegraphics[scale=0.5]{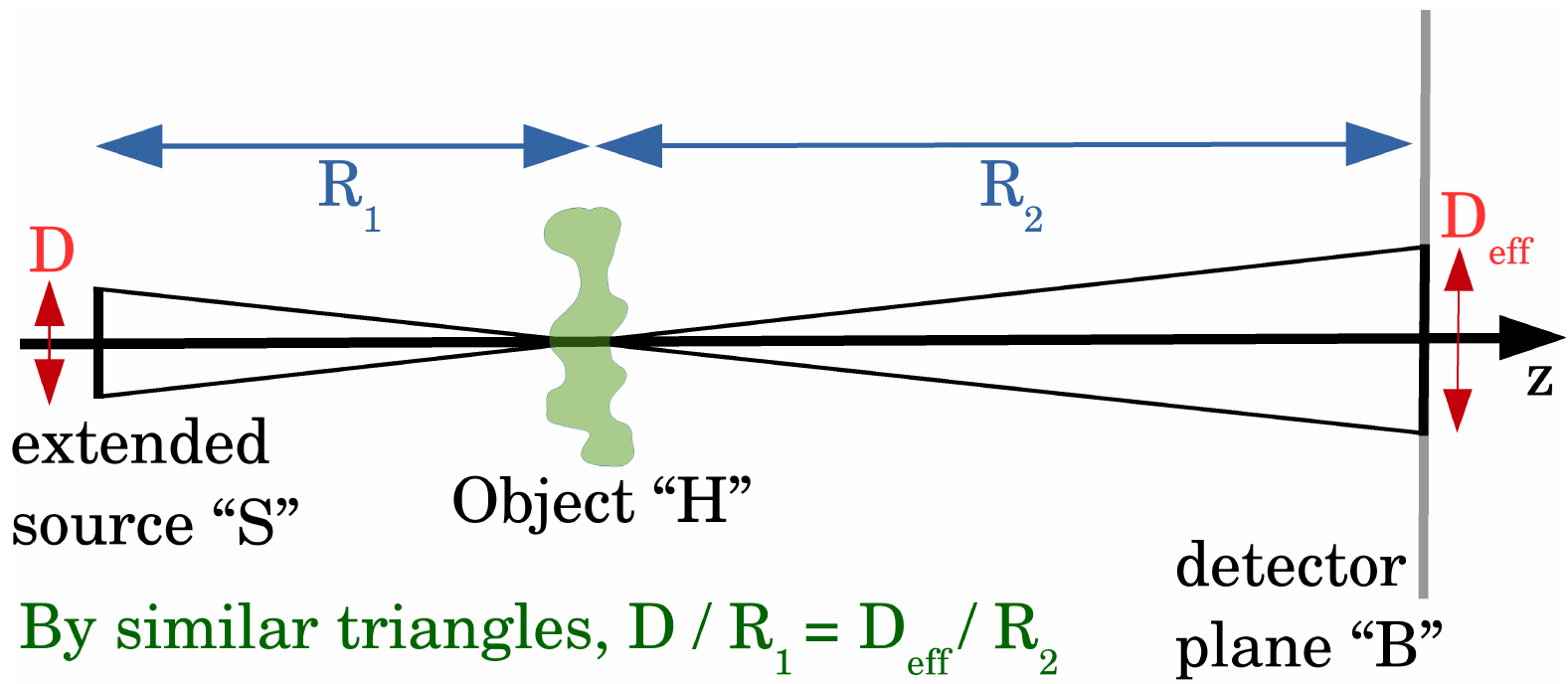}
\caption{An extended incoherent source $S$, with diameter $D$, leads to source-size blurring of spatial width $D_{\rm eff}=D R_2 / R_1$ over the detector plane $B$.}
\label{fig:Blurring}
\end{figure}

Before proceeding, we need to briefly outline the idea of {\em image blurring due to non-zero source size}. See Fig.~\ref{fig:Blurring}. The fact, that the X-ray source is not a point, will lead to some blurring of images formed using such a source.  For a so-called ``extended incoherent source'', say a planar source of diameter $D$, one can by definition consider each point on the source to be an {\em independent radiator of X-rays}.  Provided that the source is not too large, and that the assumption of independent radiators is reasonable, each of the radiators will form a separate image of the object, with images due to separate points on the source being transversely displaced from one another.  The net effect, of adding all of the slightly-displaced images formed by each point on the extended incoherent source, is to {\em blur} the resulting image obtained over the plane $B$.  The transverse length scale, over which this blurring takes place, may be obtained via the similar-triangles construction in Fig.~\ref{fig:Blurring}.  Here, we see that the transverse spatial extent $D_{\rm eff}$ of the source-size induced blurring, of the image of the object $H$ that is obtained over the detector plane $B$, is given by $D_{\rm eff}=D R_2 / R_1$.   This effect is known as ``penumbral blurring''. Note that the source-size induced blurring becomes progressively worse as $R_2$ increases, for fixed $D$ and $R_1$.

\begin{figure}
\includegraphics[scale=0.26]{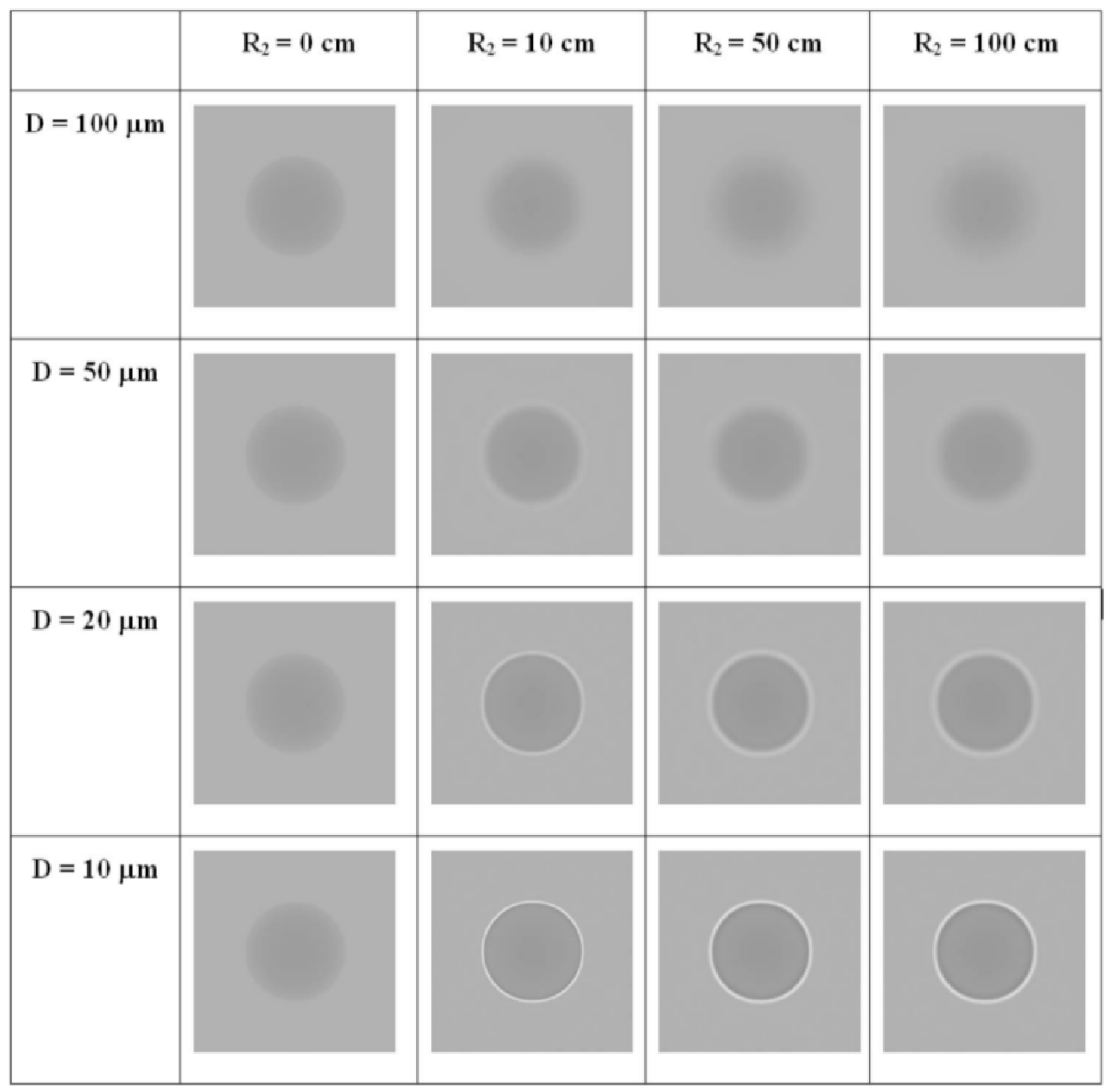}
\caption{Simulated propagation-based X-ray phase contrast images of a solid carbon sphere of diameter 0.5 mm, corresponding to X-rays with wavelength 0.5 \AA. The source-to-object distance $R_1$ is fixed at 10 cm.  The object-to-detector distance $R_2$ is increased as one moves from left to right.  The source diameter $D$ decreases as one moves from top to bottom.  Reproduced with permission from Gureyev {\em et al.}~(2009).}
\label{fig:RRR2}
\end{figure}

Following on from the above aside regarding source-size-induced blurring, we now return to the main thread of our discussion regarding propagation-based phase contrast.  To further develop the reader's intuition regarding the key features of such a mechanism for phase contrast---an image-sharpening effect which competes with penumbral image blurring---consider the simulations shown in Fig.~\ref{fig:RRR2}.  These simulations correspond to an X-ray wavelength of 0.5 \AA, and a fixed source-to-object distance $R_1$ of 10 cm.  The two variables are (i) the diameter of the source $D$, which decreases from top to bottom in the figure, and (ii) the object-to-detector propagation distance $R_2$, which increases from left to right.  The simulated sample is a solid carbon sphere with diameter 0.5 mm.  When the source has the relatively large diameter of $D=100\,\mu$m, corresponding to the top row of Fig.~\ref{fig:RRR2}, increasing the object-to-detector distance $R_2$ has the expected effect of progressively blurring the image of the sphere.  A similar trend is seen in the second row, corresponding to halving the source diameter.  All of the images in the top two rows of Fig.~\ref{fig:RRR2} may be taken as demonstrating absorption contrast alone, with a dark ``shadow'' of the carbon sphere corresponding to the absorption of X-rays that pass through the sphere.  However, in the bottom two rows of the figure, the source size is sufficiently small to have reduced the source-size-induced blurring to such a degree that propagation-based phase contrast is manifest.  Edge contrast, in the sense described earlier in this subsection, is clearly evident in both of the bottom rows, for object-to-detector propagation distances of 10 cm or greater (columns 2, 3 and 4 of the bottom two rows).  If the object-to-detector propagation distance is zero, however, one has a contact image that displays no propagation-based X-ray phase contrast (column 1).  In all of the above, one has an evident trade-off: $R_2$ must be sufficiently large to obtain propagation-induced phase contrast, while being sufficiently small for the penumbral blurring to be sufficiently mild that it does not wash out the sharpening effect of propagation-based phase contrast.

Before proceeding, we strongly recommend to readers who have not previously seen propagation-based X-ray phase contrast images, that they briefly study some of the images in one of more of the classic early papers (Snigirev {\em et al.}, 1995; Cloetens {\em et al.}, 1996; Wilkins {\em et al.}, 1996; Nugent {\em et al.}, 1996).  This will further develop the reader's intuition for the qualitative nature of such contrast, beyond what has been sketched in these notes.

\section{Practice}
\subsection{Validity of the projection approximation}\label{Sec:ValidityOfProjApprox}
The first practice tutorial discusses the limits of validity of the projection approximation, and some of the consequences of the breakdown of the projection approximation for high resolution X-ray microscopy. \\

\begin{figure}
\includegraphics[scale=0.25]{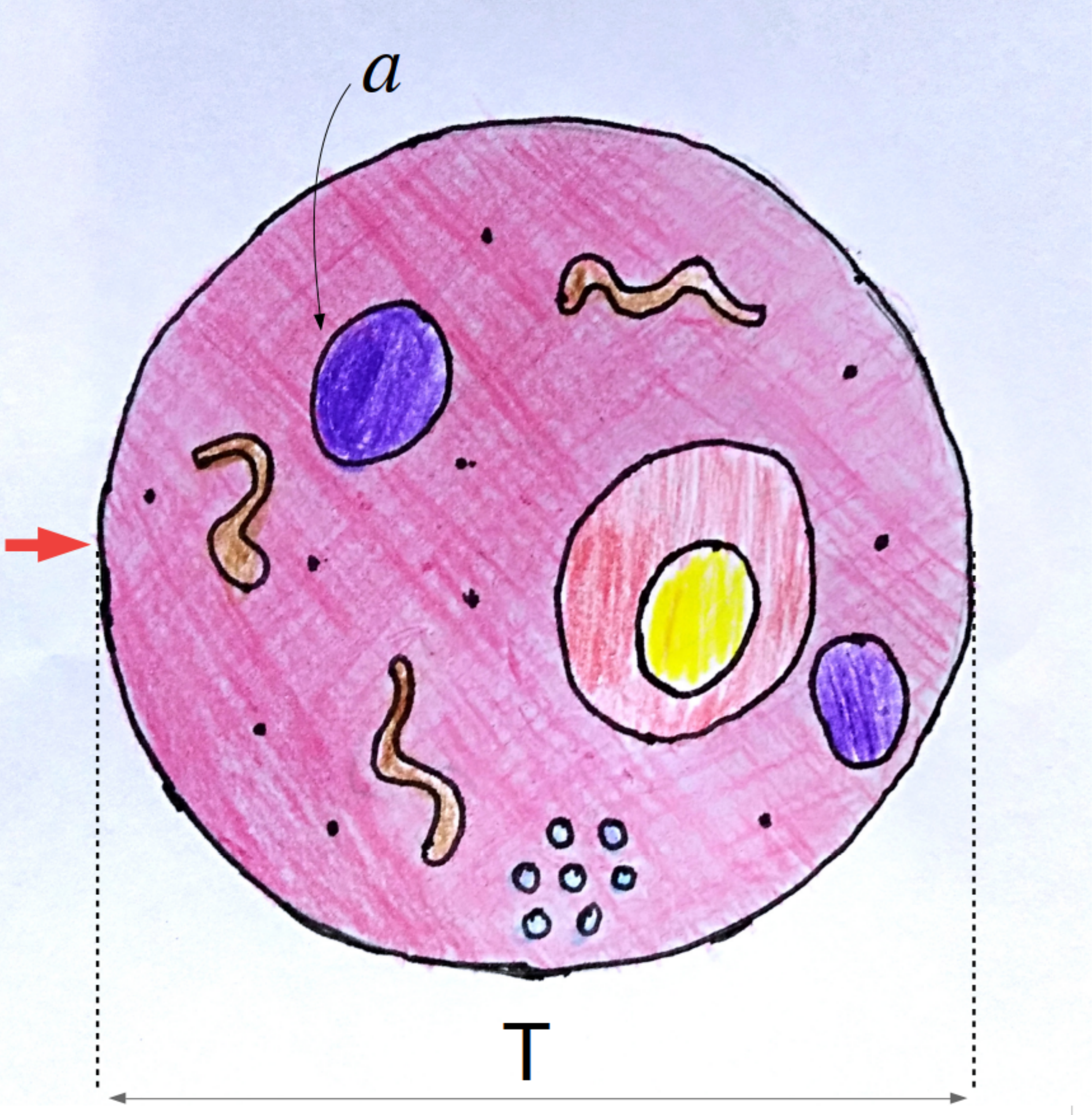}
\caption{Diagram and notations used in the discussion of the projection approximation. $T$ is the sample thickness and $a$ is the feature of interest. With some generalisation, one could assume $a$ to be the resolution of the imaging system.  Image drawn by Kristina Pelliccia.}
\label{fig:cell}
\end{figure}
 
Let us suppose we want to image a cell of thickness $T$, with the aim of resolving an organelle of size $a$ within the cell. The projection approximation in this configuration is valid if we can neglect diffraction effects within the sample, {\em i.e.}~we can assume that X-rays propagate along straight lines in the sample. Radiation of wavelength $\lambda$ scattered by the organelle will have a typical (maximum) diffraction angle of the order of
\begin{equation}
\Delta \theta = \frac{\lambda}{a}.
\label{eq:DiffractionAngle}
\end{equation}
Therefore the maximum spread of the radiation at the exit face of the sample (assuming the organelle to be close to the entrance face) will be $\Delta \theta \, T$. The projection approximation is valid if we can neglect the diffraction spread when compared to the resolution, {\em i.e.} 
\begin{equation}
\frac{\lambda}{a} \, T \ll a.
\end{equation}
The previous inequality can be redefined in terms of the Fresnel number
\begin{equation}
N_F = \frac{a^2}{\lambda T} \gg1.
\label{eq:FresnelNumber}
\end{equation}
\begin{figure}
\includegraphics[scale=0.4]{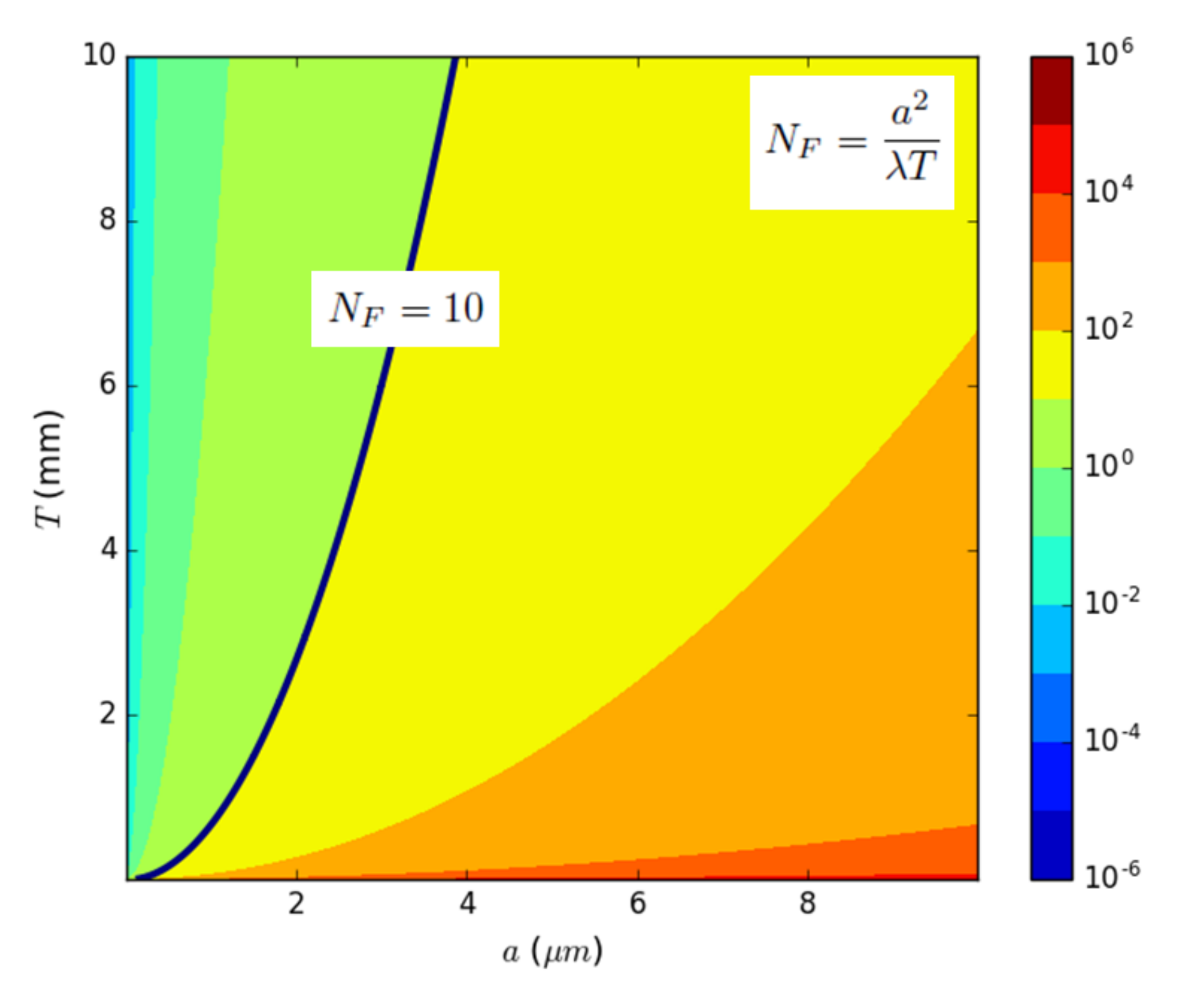}
\caption{Contour map of the Fresnel number $N_F$, as a function of resolution $a$ and sample thickness $T$, calculated at the X-ray wavelength $\lambda = 1.5$ \AA. The black line marks the contour $N_F = 10$. As a rule of thumb, a Fresnel number below this value is a situation where the projection approximation might not hold.}
\label{fig:PAContour}
\end{figure}
Figure  \ref{fig:PAContour} shows a contour map of the Fresnel number as a function of both $a$ and $T$, calculated setting the wavelength to $\lambda = 1.5$ \AA. Somewhat arbitrarily, the contour $N_F = 10$, marked with a thicker line, is chosen as a boundary for the validity of the projection approximation. The region on the right of this line (warm colours such as yellow, orange and red) is where the projection approximation is generally valid. This region corresponds to the range of values attained, for instance, by modern micro-CT (computed tomography) systems, where resolution of a few micrometers and sample thickness of a few millimetres are the state of the art.

The region on the left of the $N_F = 10$ contour (cold colours) is where the projection approximation is at risk. In this region, namely the domain of ultra high resolution X-ray microscopy systems typical of synchrotron beam-lines, the sample thickness becomes very large compared to the resolution. Admittedly, in this region lays one of the major strength of X-rays when compared with other probes for microscopy: X-rays can visualise minute details within larger samples---for instance single cells within a larger tissue---in a less invasive fashion.

Quantitative analysis at such a high resolution level however, requires one to take into account that the projection approximation may no longer hold. Let us briefly discuss two consequences of this fact. The first deals with solving the inverse problem of tomography; the second implication is relevant when modelling X-ray optical elements.

\subsection{X-ray tomography beyond the projection approximation}
One of the main consequences of the projection approximation is that the attenuation and phase shift experienced by an X-ray monochromatic beam can be expressed as line integrals, as in Eqs~\ref{eq:PhaseShiftProjectionApproximation} and \ref{eq:zz8a}. That is, X-ray tomography is based on a geometrical model of the propagation of X-rays through samples. 

Therefore, one of the most important consequences---as far as X-ray imaging is concerned---of the failure of the projection approximation, is that conventional tomography algorithms must be revisited. Specifically, the well known Fourier slice theorem is no longer valid. In this case, one must turn to what has been termed  ``diffraction tomography'', which has found wide applications in optical 3D imaging of semi-transparent samples.

The concept of diffraction tomography was first introduced by Wolf (1969). A recent review of the literature of diffraction tomography can be found in M\"uller {\em et al.}, 2016. Diffraction tomography makes use of the so-called Fourier diffraction theorem, which reduces to the Fourier slice theorem in the geometrical-optics limit (Gbur and Wolf, 2001).   

\subsection{Describing the propagation through thick samples: multi-slice approach}
\begin{figure}
\includegraphics[scale=0.5]{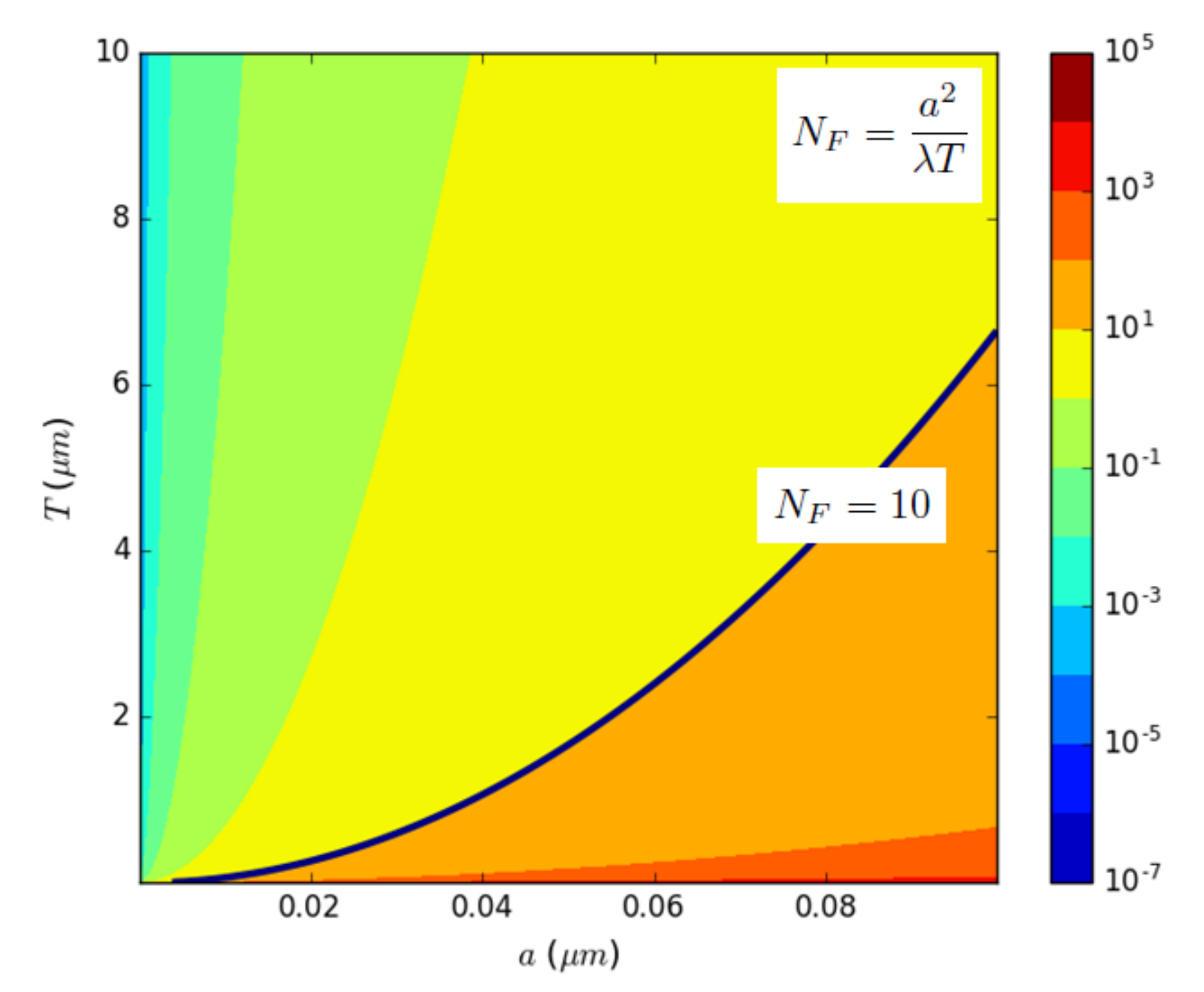}
\caption{Contour map of the Fresnel number $N_F$, as a function of resolution $a$ and sample thickness $T$, calculated at the X-ray wavelength $\lambda = 1.5$ \AA, for the high resolution case applicable to modelling X-ray diffractive optics. As before, the black line marks the contour $N_F = 10$.}
\label{fig:PAContour_hires}
\end{figure}

Simulating and modelling high resolution transmission X-ray optics, or reflective optics, is a second example of situations where the projection approximation generally does not hold. Transmission optics such as refractive X-ray lenses or Bragg--Fresnel lenses can be, to some extent, considered thin in the medium resolution range. High resolution applications however, demand extremely fine X-ray optical structures (for instance outermost zone of Fresnel or Bragg--Fresnel lenses are in the nm range). This fact can be appreciated in Fig.~\ref{fig:PAContour_hires}, which is a close-up view of the contour map in Fig.~\ref{fig:PAContour}, applicable in the region relevant to high resolution X-ray optics. \\

Modelling X-ray propagation through such elements always requires dropping the projection approximation in favour of a more accurate approach. Furthermore, conventional reflective optics must be considered ``thick'' in all cases, as obviously the beam angular deviation in reflection is always significant. In all those cases, the multi-slice approximation is a very useful approach.

Originally introduced by Cowley and Moodie (1957 and 1959) in the context of Transmission Electron Microscopy, the multi-slice approximation is being increasingly used to simulate high resolution X-ray optics and imaging. See for instance Paganin (2006), Martz {\em et al.}~(2007), D\"oring {\em et al.}~(2013) or Li {\em et al.}~(2017). Incidentally, and building upon a deliberately-provocative remark made earlier in these notes, contemporary work in X-ray multi-slice gives an excellent example of how progress in X-ray optics would be accelerated by more workers in this field being familiar with electron optics, since the multi-slice method was brought to a very high state of development by the electron-optics community, decades before the method began to be employed in earnest by the X-ray optics community.

In the multi-slice approach, the thick sample is decomposed into a number of slices along the optic axis direction. The thickness of each slice should be chosen to guarantee that such a slice can be considered optically thin. This corresponds to $N_F \gg 1$ for each individual slice. Therefore, for each slice one can assume the projection approximation to be valid.

Following Eq.~\ref{eq:zz8}, and dropping the subscript $\omega$ for clarity, the transmission function of the slice $j$ can be written as:
\begin{equation}
\mathcal{T}_{j}(x,y) = \exp \left\{ -i k \, \tilde{n_j}(x,y) \, \Delta z \right\}.
\label{eq:TranFunctSlice}
\end{equation}
In Eq.~\ref{eq:TranFunctSlice}, 
\begin{eqnarray}
\nonumber\tilde{n_j}(x,y) &=& \int_{z_{j-1}}^{z_j} \!\!\! \tilde{n}(x,y,z) dz / \Delta z \\ &\approx&  \tilde{n}(x,y, z=z_j) \label{eq:OneSliceOfMultiSlice}
\end{eqnarray}
is the complex refractive index of slice $j$, located at the longitudinal position $z=z_j$ and, with similar notation, 
\begin{equation}
\mathcal{T}_{j}(x,y) \equiv \mathcal{T}(x,y, z=z_j). 
\end{equation}
The slice thickness is 
\begin{equation}
\Delta z = z_{j}-z_{j-1}.
\end{equation}
Note that, in deriving Eq.~\ref{eq:TranFunctSlice}, by passing from the first to the second line of Eq.~\ref{eq:OneSliceOfMultiSlice}, we assumed the refractive index of each slice to be independent of $z$, within the volume occupied by the said slice.  This will be a good approximation if the slices are thin enough (compared to the length scale over which $\tilde{n}$ varies).

Under these assumptions, the propagation of the wave field to the next slice can be performed using Fresnel propagation in vacuum, using Eq.~\ref{eq:FresnelDiffraction}:
\begin{equation}
\psi_{j+1}(x,y)={\mathcal{D}}_{\Delta z} \, \left[ \psi_{j}(x,y) \mathcal{T}_{j}(x,y) \right]. 
\end{equation}

The multi-slice algorithm applies this procedure iteratively, to propagate through all slices of the sample.  The previously-cited papers of Martz {\em et al.}~(2007), D\"oring {\em et al.}~(2013) and Li {\em et al.}~(2017) give excellent examples of the application of this very powerful and general method for considering X-ray interactions with samples, in situations where the projection approximation has broken down.  For those seeking to apply the multi-slice method in an X-ray setting, much can be learned from the electron-optics text by Kirkland (2010).

\part{Elements of X-ray phase retrieval}

The second part of our notes introduces the transport-of-intensity equation, as a means for quantifying the contrast present in propagation-based X-ray phase contrast images.  We then consider generalised phase contrast X-ray imaging systems, these being an infinite variety of imaging systems that yield phase contrast in the sense that they are sensitive to the refractive (phase) effects of X-ray-transparent samples.  Finally, the inverse problem of phase retrieval (namely the decoding of X-ray phase contrast images to obtain information regarding the object that resulted in such images) is considered, and applied to both two-dimensional and three-dimensional phase-contrast X-ray imaging.

\section{Theory}

\subsection{Transport-of-intensity equation (TIE)}

Substitute Eq.~\ref{eq:MadelungDecomposition} into Eq.~\ref{eq:ParaxialEquation}, expand, cancel a common factor, and then take the imaginary part.  This gives a continuity equation expressing local conservation of optical energy (Teague, 1983; {\em cf.} Madelung, 1927), called the {\em transport of intensity equation} (TIE):
\begin{eqnarray}
-\nabla_{\perp}\cdot\left[I(x,y,z)\nabla_{\perp}\phi(x,y,z)\right]
=k\frac{\partial I(x,y,z)}{\partial z}. \label{eq:TIE}
\end{eqnarray}

Physically, this equation asserts that the divergence of the transverse Poynting vector (transverse energy-flow vector) ${\bf S} \propto I \nabla_{\perp}\phi$ governs the longitudinal rate of change of intensity.  If the divergence of the Poynting vector is positive, because the wave-field is locally behaving as an expanding wave, optical energy will be moving away from the local optic axis and so the longitudinal derivative of intensity will be negative (local defocusing; see points $J$ and $K$ in Fig.~\ref{fig:RRR1} for an example).  Conversely, if the divergence of the Poynting vector is negative, because the wave-field is locally contracting, optical energy will be moving towards the local optic axis and so the longitudinal derivative of intensity will be positive (local focusing; see points $L$ and $M$ in Fig.~\ref{fig:RRR1} for an example). Indeed, if we speak of the negative divergence ``$-\nabla_{\perp}\cdot$'' as the ``convergence'', then the TIE merely makes the intuitive statement that ``the convergence of the transverse Poynting vector is proportional to the $z$ rate of change of intensity'': thus (i) a converging wave (positive convergence or negative divergence) has a positive rate of change of intensity with respect to $z$ because optical energy is being concentrated (focused) as $z$ increases (see again the points $L$ and $M$ in Fig.~\ref{fig:RRR1}); (ii) conversely, a diverging wave (negative convergence or positive divergence) has a negative rate of change of intensity with respect to $z$ because optical energy is being rarefied (defocused) as $z$ increases (points $J$ and $K$ in Fig.~\ref{fig:RRR1}).

The above comments also pertain to the form of the TIE obtained if the finite-difference approximation
\begin{equation}
\frac{\partial I(x,y,z)}{\partial z} \approx \frac{I(x,y,z+\delta
z)-I(x,y,z)}{\delta z}
\end{equation}
\noindent is substituted into Eq.~\ref{eq:TIE}, before being solved for the propagated intensity, to give the following approximate description for propagation-based phase contrast, in the regime of sufficiently small propagation distance $\delta z$:
\begin{eqnarray}\label{eq:TIE_for_defocus}
I(x,y,z+\delta z) \approx I(x,y,z) \qquad\qquad\qquad\qquad \\ \nonumber -\frac{\delta
z}{k}\nabla_{\perp}\cdot\left[I(x,y,z)\nabla_{\perp}\phi(x,y,z)\right]. 
\end{eqnarray}

Propagation-based methods are not the only means by which phase contrast can be achieved.  {\em Many other extremely important methods exist}, including methods utilising crystals ({\em e.g.}~F\"{o}rster {\em et al.}~(1980)), diffractive imaging from far-field patterns (Miao {\em et al.}~1999), perfect gratings (Momose \textit{et al.} 2003; Weitkamp \textit{et al.} 2005; Pfeiffer \textit{et al.} 2008), random gratings (B\'{e}rujon {\em et al.}, 2012; Morgan {\em et al.}, 2012; see Zdora (2018) for a comprehensive review), edge illumination ({\em e.g.}~Diemoz {\em et al.}, 2017, and references therein), ptychography (Pfeiffer, 2018) and of course interferometry (Bonse and Hart, 1965).  Due to time limitations, these will not be reviewed here, but we note that (i) some of these methods will be briefly covered in the practice sessions, in Sec.~\ref{sec:Practice}; (ii) many of these methods can be considered to be special cases of the set of all possible linear shift invariant phase contrast imaging systems, which will be treated later in the present text. Taken together, the previously listed suite of methods forms a powerful toolbox for the X-ray imaging of samples, with each method having its particular strengths and limitations.  No method is superior to all others in all scenarios and circumstances.

\subsection{Arbitrary imaging systems}

We have already seen that the act of free-space propagation, from plane to plane, can achieve phase contrast in the sense that the propagated image (over the downstream plane, such as that given by the detector $B$ in Fig.~\ref{fig:RRR1}) has a transverse intensity distribution that depends on the transverse X-ray phase shifts is an upstream plane (such as the plane at the exit-surface of the object in Fig.~\ref{fig:RRR1}).  What happens if we generalise this propagation-based X-ray phase-contrast-imaging scenario, to a more general X-ray phase-contrast-imaging setup, by interposing an optical imaging system in between the object and the detector? 

\begin{figure}
\includegraphics[scale=0.25]{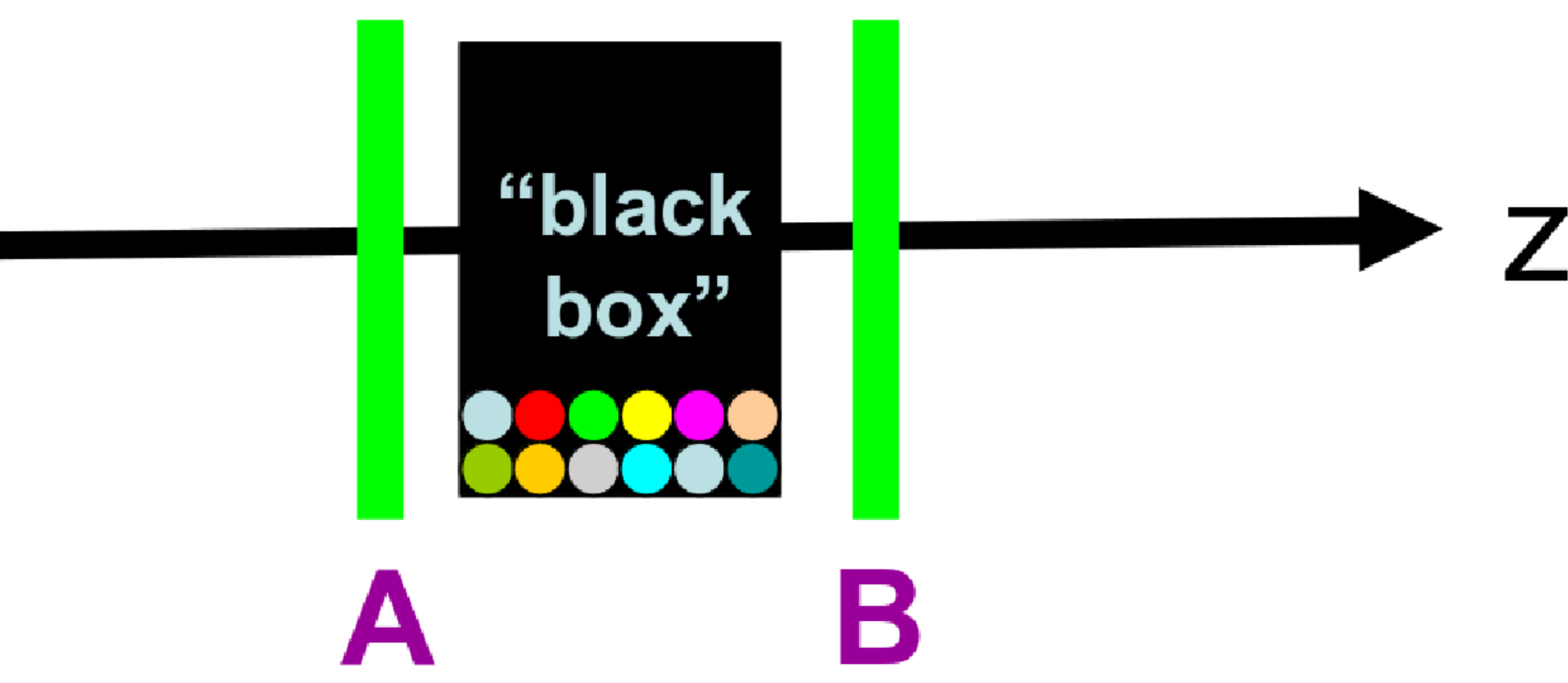}
\caption{Generalised phase-contrast imaging system.  X-rays from a source to the far left (not shown) pass through a sample (not shown), the exit surface of which is denoted $A$.  The wave-field over this plane $A$, which is assumed to be a paraxial beam travelling in the $z$ direction, is then input into an arbitrary imaging system denoted by the black box. The corresponding output complex wave-field exists over the plane $B$.  The state of the black box is schematically denoted by the coloured dials, representing the control parameters $\tau_1,\tau_2$ {\em etc}.}
\label{fig:AbbimagingSystem}
\end{figure}

Consider an arbitrary coherent X-ray imaging system that takes a two-dimensional monochromatic paraxial complex X-ray wave-field $\psi_{\rm IN}(x,y)$ as input: this corresponds to the $(x,y)$ plane labelled ``A'' in Fig.~\ref{fig:AbbimagingSystem}, which is perpendicular to the optic axis $z$.  Assume also that the state of the imaging system can be characterised by a set of real {\em control parameters} $\tau_1, \tau_2, \cdots$, with $\psi_{\rm OUT}(x,y,\tau_1,\tau_2,\cdots)$ being the corresponding complex output wave-field.

We may consider the action of this imaging system, in operator terms\footnote{For our purposes, an operator ``acts'' on a given function to give a new function.  Thus, if the operator $\mathcal{A}$ acts on the function $f$ to give a different function $g$, this would be written as $\mathcal{A}$f=g.  We follow the usual convention that each
operator acts on the element to the right of it, with the rightmost operator acting first: for example, if $\mathcal{A},\mathcal{B}$ are two operators, then $\mathcal{BA}f$ is the same as $\mathcal{B}(\mathcal{A}f)$, so that $f$ is first acted upon by $\mathcal{A}$ to give $\mathcal{A}f$, with the result being subsequently acted upon by $\mathcal{B}$ to give $\mathcal{BA}f$.}.  That is, we may consider the imaging system to be described by an operator ${\mathcal{D}}(\tau_1,\tau_2,\cdots)$ that acts on the input field to give the output field.  This may be written in the following way:
\begin{eqnarray}
\psi_{\rm OUT}(x,y,\tau_1,\tau_2,\cdots)
={\mathcal{D}}(\tau_1,\tau_2,\cdots)\psi_{\rm IN}(x,y).
\label{eq:ArbitraryCoherentImagingSystem}
\end{eqnarray}

At this stage our imaging system has a very high degree of generality: its arbitrariness is limited only by the implicit assumptions associated with a forward-propagating monochromatic scalar input being mapped to a forward-propagating monochromatic scalar output that has the same energy\footnote{These implicit assumptions include the imaging system being time-independent, elastically scattering and non back-scattering.}.

\subsection{Arbitrary linear imaging systems}

Make the further assumption that the imaging system is {\em linear}, {\em i.e.}~that the output field is a linear function of the input field.  Stated differently, we are here assuming the superposition principle to hold: if the input field is given by the sum of two particular input fields $\alpha\psi_{\rm IN}^{(1)}(x,y)+\beta\psi_{\rm IN}^{(2)}(x,y)$, where $\alpha$ and $\beta$ are arbitrary complex weighting coefficients, then the output field will (by assumption) always be equal to the sum of corresponding outputs, {\em i.e.}~$\mathcal{D}[\alpha\psi_{\rm IN}^{(1)}(x,y)+\beta\psi_{\rm IN}^{(2)}(x,y)]=\alpha\mathcal{D}\psi_{\rm IN}^{(1)}(x,y)+\beta\mathcal{D}\psi_{\rm IN}^{(2)}(x,y)+\gamma$. The complex constant $\gamma$, while consistent with the assumption of linearity, will be set to zero since it is natural to assume that a zero input field corresponds to a zero output field.  Assume further that any magnification, rotation and shear is taken into account by appropriate choice of coordinates for the plane occupied by the output wave-field.  The action of the imaging system can then be described by the following linear integral transform\footnote{An integral transform is an integral that transforms one function into another.  A {\em linear} integral transform is an integral that (i) transforms one function into another, and (ii) has the property of linearity.  The linearity property, by definition, requires the linear integral transform of a sum of two functions, to be equal to the sum of the corresponding transforms.  Example: For a function $f(x)$ of one variable $x$, an arbitrary linear integral transform could be written as $g(x)=\int f(x') K(x,x') dx' + L(x)$, where $K(x,x')$ and $L(x)$ are arbitrary functions.  In the main text, we use linear integral transforms to represent the action of linear imaging systems.  Here, the linear integral transform serves to change (transform!) the field input into the imaging system, into the field that is output by the imaging system.  Finally, we note that: (i) The function $K(x,x')$ is often called the {\em kernel} of the linear integral transform; (ii) if one can assume that a zero input gives a zero output, then $L(x)=0$.}, which may be viewed as a continuous form of matrix multiplication:  
\begin{eqnarray}\label{eq:ArbitraryLinearCoherentImagingSystem}  
\psi_{\rm
OUT}(x,y,\tau_1,\tau_2,\cdots) \qquad\qquad\qquad\qquad\qquad\qquad \\ \nonumber =\iint dx' dy'
G(x,y,x',y',\tau_1,\tau_2,\cdots)\psi_{\rm IN}(x',y').
\end{eqnarray}

The kernel of the above linear integral transform has been denoted by $G$, since it is a Green function.  It may also be interpreted as a generalised Huygens wavelet.  

To see this latter point, choose the special case 
\begin{equation}
\psi_{\rm IN}(x',y')=\delta(x'-x_0,y'-y_0)
\end{equation}
in the above expression, where $\delta(x,y)$ is a two-dimensional Dirac delta, corresponding to a single point 
\begin{equation}
(x,y)=(x_0,y_0)
\end{equation}
being illuminated in the input plane of the imaging system.  Via the sifting property of the Dirac delta\footnote{Here and elsewhere, the reader is assumed to be familiar with the basics of Fourier analysis in an optics context.  Such basics include the sifting property of the Dirac delta, the concept of convolution, the convolution theorem, and the Fourier derivative theorem.  See {\em e.g.}~Appendix A in Paganin (2006), for an overview of these basics that employs a notation consistent with these notes.}, Eq.~\ref{eq:ArbitraryLinearCoherentImagingSystem} gives the associated output field as $G(x,y,x_0,y_0,\tau_1,\tau_2,\cdots)$.  Therefore $G(x,y,x_0,y_0,\tau_1,\tau_2,\cdots)$ is the output field as a function of $x$ and $y$ coordinates in the output plane, which would be obtained if a unit-strength point source were to be located at position $(x_0,y_0)$ in the input plane, and the imaging system interposed between input and output plane were to have the state characterised by the particular control parameters $\tau_1,\tau_2,\cdots$.  Hence $G(x,y,x_0,y_0,\tau_1,\tau_2,\cdots)$ is indeed a generalised Huygens-type wavelet, with the form of the wavelet depending on both the state of the imaging system and on the position $(x_0,y_0)$ of the input ``pinpoint of X-ray light''.  

We close this sub-section by reversing the chain of logic that is given above, so as to physically motivate the writing down of Eq.~\ref{eq:ArbitraryLinearCoherentImagingSystem} for an arbitrary linear imaging system.  We characterise such an imaging system by the fact that, if the input is a ``pinpoint of X-ray light'' $\delta(x-x',y-y')$ at some point $(x',y')$ in the entrance plane $A$ of Fig.~\ref{fig:AbbimagingSystem}, then the corresponding output field---considered as a function of coordinates $(x,y)$ over the output plane $B$---will be given by $G(x,y,x',y',\tau_1,\tau_2,\cdots)$.  In this expression for the output field $G$, the coordinates $(x',y')$ of the input ``pinpoint of X-ray light'' are considered to be fixed, with the parameters $\tau_1,\tau_2,\cdots$ describing the state of the imaging system also being fixed. To proceed further, we can used the sifting property of the Dirac delta to decompose an {\em arbitrary} input field $\psi_{\rm IN}(x,y)$ as a superposition (described by the continuous sum, namely the integral sign below) of X-ray pinpoints of light, each such pinpoint having the form $\delta(x-x',y-y')$, so that:
\begin{equation}
\psi_{\rm IN}(x,y)=\iint dx' dy' \psi_{\rm IN}(x',y') \delta(x-x',y-y').
\end{equation}
In order to map inputs to outputs, namely to convert $\psi_{\rm IN}(x,y)$ in the above integral (superposition of pinpoint inputs, each of which have the form $\delta(x-x',y-y')$ multiplied by a weighting coefficient $\psi_{\rm IN}(x',y')$) into $\psi_{\rm OUT}(x,y)$, we need only replace each of the pinpoint inputs $\delta(x-x',y-y')$ under the integral sign, with its corresponding output $G(x,y,x',y',\tau_1,\tau_2,\cdots)$.  This direct employment of the superposition principle---which is justified on account of our key assumption that the imaging system is {\em linear}---leads directly to Eq.~\ref{eq:ArbitraryLinearCoherentImagingSystem}.

\subsection{Arbitrary linear shift-invariant imaging systems}

We specialise still further, by assuming the linear imaging system to be {\em shift invariant}.  This augments the previous assumptions, with the additional assumption that, if there is a transverse shift of the input wave-field, this merely serves to transversely shift the output wave-field. Such an assumption cannot hold for arbitrarily large transverse shifts, but is often approximately true for a sufficiently small range of transverse shifts in the vicinity of the centre of the field of view of a coherent linear imaging system.  The assumption of (transverse) shift invariance implies that Eq.~\ref{eq:ArbitraryLinearCoherentImagingSystem} may be simplified to:    
\begin{eqnarray}\label{eq:ArbitraryShiftInvariantLinearCoherentImagingSystem1}
\qquad \psi_{\rm
OUT}(x,y,\tau_1,\tau_2,\cdots) \qquad\qquad\qquad\qquad\qquad\qquad\qquad \\ \nonumber =\iint dx' dy'
G(x-x',y-y',\tau_1,\tau_2,\cdots)\psi_{\rm IN}(x',y').
\end{eqnarray}

This will be recognised as a two-dimensional convolution (folding, Faltung) integral, and hence may be more compactly written as:
\begin{eqnarray}\label{eq:ArbitraryShiftInvariantLinearCoherentImagingSystem2}
\psi_{\rm
OUT}(x,y,\tau_1,\tau_2,\cdots) \qquad\qquad\qquad\qquad\qquad\qquad \\ \nonumber = \psi_{\rm IN}(x,y) \star G(x,y,\tau_1,\tau_2,\cdots),
\end{eqnarray}
\noindent where $\star$ denotes two-dimensional convolution.  

A very rich variety of imaging systems in coherent X-ray optics may be described using the formalism based on Eq.~\ref{eq:ArbitraryShiftInvariantLinearCoherentImagingSystem1}, including propagation-based X-ray phase contrast, analyser-crystal-based phase contrast, imaging/microscopy using compound refractive lenses, imaging/microscopy using Fresnel zone plates, inline holography, off-axis holography, Zernike phase contrast imaging, imaging/microscopy using Kirkpatrick--Baez mirrors, interferometry, grating-based X-ray imaging, speckle-tracking X-ray imaging and various forms of imaging system that perform optical encryption.   However, systems such as reflective optics and X-ray wave-guides, where multi-slice is required to describe passage of X-rays through optical elements, need the more general form given by  Eq.~\ref{eq:ArbitraryLinearCoherentImagingSystem}.

\subsection{Transfer function formalism}

Fourier transform\footnote{We use the Fourier-transform convention from Appendix A of Paganin (2006). In one spatial dimension, the Fourier transform $\mathcal{F}$ of a function $g(x)$ is denoted by $\mathcal{F}[g(x)]\equiv\breve{g}(k_x)$, where $k_x$ is the Fourier coordinate corresponding to $x$, and $\breve{g}(k_x)=(1/\sqrt{2\pi})\int_{-\infty}^{\infty} g(x) \exp[-ik_xx]dx$, with $g(x)=(1/\sqrt{2\pi})\int_{-\infty}^{\infty} \breve{g}(k_x)\exp[ik_xx]dk_x$ denoting the corresponding inverse Fourier transform.  In two dimensions, and in an obvious extension of the notation, the forward Fourier transform becomes $\breve{g}(k_x,k_y)=(1/(2\pi))\iint_{-\infty}^{\infty} g(x,y) \exp[-i(k_xx+k_yy)]dxdy$, and the inverse transform becomes $g(x,y)=(1/(2\pi))\iint_{-\infty}^{\infty} \breve{g}(k_x,k_y)\exp[i(k_xx+k_yy)]dk_xdk_y$.} both sides of Eq.~\ref{eq:ArbitraryShiftInvariantLinearCoherentImagingSystem2} with respect to $x$ and $y$, indicated by the operator $\mathcal F$.  Invoke the convolution theorem of Fourier analysis, to convert convolution to multiplication. The inverse Fourier transform of the resulting expression is the following operator-type description of the action of the imaging system:
\begin{eqnarray}
\psi_{\rm
OUT}(x,y,\tau_1,\tau_2,\cdots) = {{\mathcal{D}}}(\tau_1,\tau_2,\cdots) \psi_{\rm IN}(x,y)
\label{eq:ArbitraryShiftInvariantLinearCoherentImagingSystem3}
\end{eqnarray}
\noindent where the Fourier transform of our Huygens-type wavelet has been termed the {\em transfer function}: 
\begin{eqnarray}
T(k_x,k_y,\tau_1,\tau_2,\cdots)\equiv 2\pi\mathcal{F}
[G(x,y,\tau_1,\tau_2,\cdots)],
\label{eq:ArbitraryShiftInvariantLinearCoherentImagingSystem4}
\end{eqnarray}
\noindent $(k_x,k_y)$ denotes Fourier-space (spatial frequency)  coordinates corresponding to real-space coordinates $(x,y)$, and the generalised diffraction operator quantifying our imaging system is the following Fourier-space filtration:
\begin{eqnarray} {{\mathcal{D}}}(\tau_1,\tau_2,\cdots)=
\mathcal{F}^{-1}T(k_x,k_y,\tau_1,\tau_2,\cdots)\mathcal{F}.
\label{eq:ArbitraryShiftInvariantLinearCoherentImagingSystem5}
\end{eqnarray}

\noindent In the above, it is important to recall that all operators act from right to left: {\em i.e.}~if the operator ${{\mathcal{D}}}(\tau_1,\tau_2,\cdots)$ is applied to an input field, that input field is first acted upon by the Fourier transform $\mathcal F$, then multiplied by the transfer function ${T}$, and then inverse Fourier transformed.  We previously stated this as, ``We follow the usual convention that each operator acts on the element to the right of it, with the rightmost operator acting first.''

In words, Eqs~\ref{eq:ArbitraryShiftInvariantLinearCoherentImagingSystem3} and \ref{eq:ArbitraryShiftInvariantLinearCoherentImagingSystem5} state the following: In order to map the input field $\psi_{\rm IN}(x,y)$ to the corresponding field $\psi_{\rm OUT}(x,y)$ output by a linear shift-invariant imaging system, a sequence of three steps may be used:
\begin{enumerate}

    \item Apply the Fourier transform operator $\mathcal F$ to the input field;

    \item Multiply the resulting object, which will be a function of the Fourier coordinates $(k_x,k_y)$, by the transfer function $T(k_x,k_y,\tau_1,\tau_2,\cdots)$ corresponding to the linear shift-invariant imaging system being in a state described by the control parameters $(\tau_1,\tau_2,\cdots)$;

    \item Apply the inverse Fourier transform operator.

\end{enumerate}
This verbal description may be considered as pseudo code for a computational simulation of a linear shift-invariant imaging system; the resulting computer codes are typically rendered extremely efficient by the use of the fast Fourier transform (FFT) to implement both the forward and inverse Fourier transform operators.  From a more physical perspective: Step 1 is a {\em decomposition} of the input field into its constituent plane-wave components (Fourier components), Step 2 is a {\em filtration} of these plane-wave components in which each such plane-wave component is weighted by a different multiplicative factor that is given by the transfer function $T$, and Step 3 is a {\em synthesis} in which all of the resulting weighted plane waves are added up to give the output field. For more on the synthesis--decomposition concept in optics, we refer the reader to Gureyev {\em et al.}~(2018).     

An important special case of a linear shift-invariant imaging system, is the previously considered case of free-space propagation through vacuum by a distance $\Delta$, in which case we write  ${{\mathcal{D}}}(\tau_1,\tau_2,\cdots)\longrightarrow {\mathcal{D}}_{\Delta}$, with
\begin{equation}\label{eq:DiffractionOperatorFresnelDiffraction}
   {\mathcal{D}}_{\Delta}=\exp(ik\Delta)\mathcal{F}^{-1}
   \exp\left[\frac{-i\Delta(k_x^2+k_y^2)}{2k}\right]\mathcal{F}.
\end{equation}
This is a two-Fourier-transform version of the Fresnel diffraction integral.  For more detail on this connection, we refer the reader to Paganin (2006).

A second important special case corresponds to analyser-based X-ray phase contrast, where the X-ray field transmitted through a sample is reflected from the surface of a near-perfect crystal before having its intensity registered by a position-sensitive detector.  In this case, upon suitable rotation of the $(x,y)$ coordinates, 
\begin{eqnarray} {{\mathcal{D}}}(\tau_1,\tau_2,\cdots)\longrightarrow \mathcal{A} = \mathcal{F}^{-1} A(k_x) \mathcal{F},
\label{eq:ArbitraryShiftInvariantLinearCoherentImagingSystem6}
\end{eqnarray}
\noindent where the analyser-crystal transfer function $A(k_x)$ is a polarisation-dependent function of $k_x$ whose exact form is not needed here ({\em cf.}~Paganin {\em et al.}, 2004a,b).

\subsection{Phase contrast}

The squared magnitude of the input--output equation
\begin{equation}
\psi_{\rm
OUT}(x,y,\tau_1,\tau_2,\cdots)= {\mathcal D}(\tau_1, \tau_2, \cdots) \psi_{\rm IN}(x,y)
\end{equation}
gives the intensity, output by our shift-invariant linear imaging system, as:
\begin{eqnarray}\label{eq:IntensityTransfer}
I_{\rm
OUT}(x,y,\tau_1,\tau_2,\cdots)= |{\mathcal D}(\tau_1, \tau_2, \cdots) \psi_{\rm IN}(x,y)|^2. \quad
\end{eqnarray}

Evidently---and with the important exception of the ``perfect imaging system'' case where $\mathcal{D}$ is equal to unity---the {\em output intensity typically depends upon both the intensity and phase of the input}, since the right side of Eq.~\ref{eq:IntensityTransfer} will typically {\em couple the phase of the input field to the intensity of the output field}.  Any state $(\tau_1,\tau_2,\cdots)$ of the imaging system, which generates an output intensity that is influenced by the input phase, is said to exhibit {\em phase contrast}. Again, most states of an {\em imperfect} imaging system described by the operator $\mathcal{D}\ne 1$ will yield both intensity contrast and phase contrast. 

Let us re-iterate a most important point.  If we define a ``perfect'' imaging system as one which perfectly reproduces the input field, up to magnification, then such a system will have 
\begin{equation}
{\mathcal D}(\tau_1, \tau_2, \cdots)\rightarrow 1.
\end{equation}
Equation~\ref{eq:IntensityTransfer} reduces to 
\begin{equation}
I_{\rm
OUT}(x,y,\tau_1,\tau_2,\cdots)=I_{\rm IN}(x,y).  
\end{equation}
Therefore, {\em an imaging system which is perfect at the field level, in the sense that the diffraction operator that maps input field to output field is given by} $\mathcal{D}=1${\em , yields no phase contrast.}  This trivial statement may be compared with the rather important statement that {\em imperfect (aberrated) imaging systems typically {\bf do} exhibit phase contrast}.  See {\em e.g.}~Paganin and Gureyev (2008) and Paganin {\em et al.}~(2018) for further information, regarding the nature of the phase contrast that may be associated with arbitrary linear shift-invariant imaging systems.

\subsection{Forward and inverse problems}

So-called {\em forward problems}, in physics, seek to determine effects from causes.  Examples of such forward problems include: 

\begin{enumerate}

\item Solving the Schr\"{o}dinger equation of non-relativistic quantum mechanics, to determine the allowed energy levels of a hydrogen atom;

\item Determining the spectrum of different sound pitches that would be created if a guitar string of a given length and tension {\em etc.}~were to be plucked at a particular position; 

\item Using the transfer-function formalism to calculate the intensity distribution of a propagation-based phase contrast image, for a specified sample with known three-dimensional complex refractive index, under the projection approximation, for known experimental parameters such as X-ray wavelength, source-to-detector distance {\em etc}.

\end{enumerate}

{\em Inverse problems}, on the other hand, seek to determine causes from effects. Examples include: 

\begin{enumerate}

\item Schr\"{o}dinger's inferring of his famous equation, based on data available at the time, such as the measured energy levels of the hydrogen atom; 

\item Determining the position at which a guitar string of known length is plucked, given a measurement of the spectrum of different sound pitches created by the plucked string; 

\item Determining both the magnitude and the phase of the projected complex refractive index created by a sample, under the projection approximation, for known experimental parameters such as X-ray wavelength, source-to-detector distance {\em etc.}, and a known  propagation-based phase contrast intensity image.

\end{enumerate}

If the underlying fundamental physics equations are known, and enough reasonable initial data is specified, the forward problems of classical physics are typically soluble. This broad statement is based on the fact that, in performing an experiment to model a given classical-physics scenario, {\em nature always chooses a ``solution''---namely the actual physical state for a classical system at a given specified time in its future---for a specified starting state of the system}\footnote{This solution may not necessarily be uniquely obtained from the starting point ({\em e.g.}~in dissipative systems with a point-like attractor, whereby a family of state-space trajectories may converge upon a single point in state space; see Ruelle 1989), and it may exhibit sensitive dependence upon initial conditions ({\em e.g.}~in non-dissipative chaotic systems with strange attractors; again see Ruelle 1989), but such subtleties do not change the fact that, classically speaking, ``nature always chooses a solution''.  Moreover, if one's systems of equations, which model a given scenario in the physical world, should evolve to states that are {\em singular} ({\em e.g.}~the infinite energy densities associated with ray caustics in geometric optics), then this lets one know that a more general theory is needed ({\em e.g.}~wave optics, which smooths out the infinities predicted by crossing rays in geometric optics; see Berry \& Upstill 1980, Berry 1998, and Paganin 2006); again, the existence of singularities in one's physical model does not contradict the earlier statement regarding nature always finding a solution.  Also, there may be the more subtle problem that, for a given system of equations, it may not be rigorously known whether solutions {\em to the equations as posed} even exist for certain specified classes of initial condition ({\em e.g.}~such questions remain outstanding for the Navier--Stokes equations of classical fluid mechanics; see Kreiss \& Lorenz 1989); again, such interesting subtleties do not contradict our earlier statement.}.        

Inverse problems are harder, in general, than their associated forward problems.  Solutions to specified inverse problems do not necessarily exist; even if they do exist, they may not be unique; even if a unique solution exists, it may not be stable with respect to perturbations in the data due to realistic amounts of experimental noise, and other imperfections present in any real experiment.  If an inverse problem is indeed such that there exists a unique solution that is stable with respect to perturbations in the input data, it is said to be {\em well posed in the sense of Hadamard} (Hadamard, 1923; Kress, 1984).  While such a property is desirable from both an analytic and aesthetic perspective, the class of inverse problems that scientists and engineers may wish to solve, is rather broader than the class of inverse problems that are well posed in the sense of Hadamard.  In this latter context, various forms of optimisation method are very powerful, although a treatment of such methods is beyond the scope of these notes.

\subsection{Two inverse problems}

We open this sub-section by revising what we have learned so far regarding the forward problem of imaging using generalised shift-invariant linear (phase contrast) imaging systems.  We separately consider the forward and inverse problems at the levels of (i) fields, (ii) intensities.  Note that the former problem is somewhat idealised, since complex X-ray wave-fields are not measured directly. Rather, it is {\em time-averaged intensities} that are directly measured by X-ray detectors, with the time average being taken over the acquisition time of the detector.

(i) At the field level for an arbitrary linear shift invariant imaging system, we learned that the input field may be related to the output via Eq.~\ref{eq:ArbitraryShiftInvariantLinearCoherentImagingSystem3}, with the input-to-output operator $\mathcal{D}(\tau_1,\tau_2,\cdots)$ given by Eq.~\ref{eq:ArbitraryShiftInvariantLinearCoherentImagingSystem5}.  The associated inverse problem, namely the determination of the input field given the output field, is solved by:
\begin{eqnarray}\label{eq:TransferFunctionInverseProblem}
\psi_{\rm
IN}(x,y)= \mathcal{F}^{-1} \frac{1}{T(k_x,k_y,\tau_1,\tau_2,\cdots)}\mathcal{F} \\ \nonumber\times\psi_{\rm OUT}(x,y,\tau_1,\tau_2,\cdots). 
\end{eqnarray}

Often there are division-by-zero issues associated with spatial frequencies $(k_x,k_y)$ at which the transfer function $T(k_x,k_y,\tau_1,\tau_2,\cdots)$ vanishes.  This amounts to information loss in the forward problem, leading to instability in the associated inverse problem.  Sometimes one can ``regularise'' the above expression by replacing $1/T$ with $T^*/(|T|^2+\aleph)$ where $\aleph$ is a small positive real number.  A more sophisticated solution is to consider several outputs associated with $N>1$ different states of the imaging system, leading to the following solution to the field-level inverse problem (Schiske 1968; Paganin {\em et al.} 2004c):
\begin{eqnarray}\label{eq:TransferFunctionInverseProblem2}
\psi_{\rm
IN}(x,y)= \mathcal{F}^{-1}\sum_{j=1}^{j=N} \frac{T_j^*(k_x,k_y)}{\sum_{p=1}^{p=N} |T_p(k_x,k_y)|^2}\mathcal{F} \quad \\ \nonumber\times\psi^{(j)}_{\rm OUT}(x,y). 
\end{eqnarray}
\noindent Here, $T_j(k_x,k_y)$ denotes the transfer function associated with the $jth$ state of the imaging system, and $\psi^{(j)}_{\rm OUT}(x,y)$ denotes the corresponding output.  The above expression will have no division-by-zero issues if $\sum_{p=1}^{p=N} |T_p(k_x,k_y)|^2$ is non-zero at every spatial frequency $(k_x,k_y)$.  If division-by-zero issues remain, one can always regularise the above expression, or increase the number of different states of the imaging system that is utilised.

(ii)  The inverse problem of phase retrieval, or more properly of phase--amplitude retrieval, seeks to {\em reconstruct both the intensity and phase of the input field, given only the intensity of the output field} corresponding to one or more states $(\tau_1,\tau_2,\cdots)$ of the imaging system.  This problem is vastly more difficult than the previously-considered field-level inverse problem.  Indeed, no closed form solution exists in general, to the phase--amplitude retrieval problem.  Note the evident parallels with the concept of inline holography as conceived by Gabor (1948), in which imaging is viewed as a two-step process: data recording, followed by reconstruction.  The ``holographic'' spirit of this latter point implicitly runs through many of our subsequent discussions regarding phase retrieval.  Before proceeding, however, we make the following general remark, which again has parallels with holography: Since imperfect shift-invariant aberrated imaging systems typically yield measurable output images $I_{\textrm{OUT}}(x,y)$ that are affected by the phase of the input complex field $\psi_{\textrm{IN}}(x,y)$, the output intensity may be viewed as containing {\em encrypted or encoded information regarding the phase of the input field}.  Under this view, the phase-retrieval problem corresponds to seeking a means to {\em decyrpt or decode} one or more measured output-intensity maps, so as to infer the phase distribution (or, more generally, both the phase and the amplitude/intensity) of the input field.           

\subsection{Transport-of-intensity phase retrieval}

For propagation based X-ray phase contrast imaging of a {\em single-material object} with projected thickness $T(x,y)$ that is normally illuminated by plane waves of uniform intensity $I_0$, under the projection approximation, it is evident from Eqs~\ref{eq:PhaseShiftProjectionApproximation} and \ref{eq:zz8a} that both the phase and the amplitude at the exit surface of the object may be obtained from $T(x,y)$\footnote{For a single-material object illuminated by normally incident plane waves of uniform intensity $I_0$, the phase shift in Eq.~\ref{eq:PhaseShiftProjectionApproximation} becomes $\Delta\phi(x,y)=-k \, \delta_{\omega} T(x,y)$, and the absorption-contrast intensity map in Eq.~\ref{eq:zz8a} becomes $I(x,y,z=z_0)=I_0 \exp[-\mu_{\omega} T(x,y)]$.  Here, $T(x,y)$ is the projected thickness of the single-material sample, and the wavenumber $k$ is equal to $2\pi$ divided by the X-ray wavelength $\lambda$.}.  This opens the logical possibility that the projected thickness of the single-material sample may be obtained from a single propagation-based phase contrast image $I(x,y,z=\Delta)$, obtained at a distance $\Delta$ downstream of the object that is sufficiently small for the Fresnel number\footnote{Here, the Fresnel number is as defined in Eq.~\ref{eq:FresnelNumber}, but with the important difference that the $T$ in the denominator is replaced by the object-to-detector propagation distance $\Delta$.} $N_F$ to be much greater than unity (this corresponds to the ``single edge fringe'' regime exemplified by the propagation-based phase contrast images in the bottom right of Fig.~\ref{fig:RRR2}).  With the previously mentioned approximations, but no further approximations of any kind, the transport-of-intensity equation  (Eq.~\ref{eq:TIE}) may be solved exactly, to give the projected thickness of the sample from a single propagation-based phase contrast image (Paganin {\em et al.}, 2002):
\begin{equation}\label{eq:SPEX}
T(x,y)=-\frac{1}{\mu}\log_e\left(\mathcal{F}^{-1}
\left\{\frac{\mathcal{F}\left[I(x,y,z=\Delta)\right]/I_0}
{1+(\delta\Delta/\mu)(k_x^2+k_y^2)}\right\}
\right).
\end{equation}

The above algorithm has been widely utilised, and is now known as ``Paganin's algorithm'' or ``Paganin's method''.  Its advantages, bought at the price of the previously stated strong assumptions, include simplicity, speed, very significant noise robustness and the ability to process time-dependent objects frame-by-frame.  While the method provides quantitative results when its key assumptions are sufficiently well met, qualitative reconstructions obtained under a broader set of conditions are often of utility where non-quantitative morphological information is sufficient.

A variant of the Paganin algorithm has been developed for analyser based phase contrast imaging and other phase contrast imaging systems that yield first-derivative phase contrast (Paganin {\em et al.}, 2004b).  Another variant has has been developed for phase contrast imaging systems that simultaneously yield both first-derivative and second-derivative phase contrast (Pavlov {\em et al.}, 2004 and 2005). 

\subsection{The inverse problem of tomography}

Suppose that a static non-magnetic three-dimensional object is placed upon a spindle about which the object can be rotated through a set of azimuthal angles $\varphi$ which are (say) equally spaced throughout the interval from 0 to $\pi$ radians.  Suppose further that the sample is normally illuminated with uniform intensity monochromatic scalar X-ray plane waves, and that all of the assumptions needed for the projection approximation are valid.

As we have previously learned in our discussions relating to the projection approximation, both the phase and the logarithm of the intensity, of the exit surface wavefield for each orientation, may be obtained via a simple linear projection of the complex refractive index (see Eqs~\ref{eq:PhaseShiftProjectionApproximation} and \ref{eq:zz8a_log_form} respectively).  This process may be inverted in the process of tomography, with the imaginary part of the three-dimensional complex refractive index being obtainable from measurements of the logarithm of the exit-surface intensity over the set of azimuthal angles of the spindle.  Similarly, if a suitable phase retrieval can be performed for each orientation of the object, the recovered set of two-dimensional phase maps may be inverted to give the real part of the complex refractive index of the sample.

Note that when the Paganin method is utilised in a {\em tomographic} context, its domain of utility broadens since many objects may be viewed as {\em locally} composed of a single material of interest, in three spatial dimensions, that cannot be described as composed of a single material in projection (Beltran {\em et al.}, 2010 and 2011).  In such a tomographic setting, the algorithm is sufficiently robust with respect to noise that Beltran {\em et al.}~(2010, 2011) noted it could exhibit signal-to-noise ratio (SNR) boosts of up to 85 (resp. 200).  More recent studies have shown that this SNR boost has, as an approximate upper limit, 0.3 $\delta_{\omega}/\beta_{\omega}$ if Poisson statistics are assumed (Nesterets and Gureyev, 2014; Gureyev {\em et al.}, 2014).  Interestingly, this boost in SNR can be even more marked for very small  exposure times (Kitchen {\em et al.}, 2017).  Since signal-to-noise varies with the square root of dose, this SNR-boost implies that reduction in dose of a factor of $300^2 =90,000$ is possible, at least in principle, when the Paganin method it applied to tomography (Kitchen {\em et al.}, 2017).  This fact is of importance in dose-sensitive applications of the method ({\em e.g.} to biomedical imaging), as well as time-sensitive applications where imaging speed is an issue. Largely on account of its SNR-boosting properties, new applications of the Paganin method are regularly published in a variety of fields; we refer the reader to

\bigskip

\url{https://bit.ly/2FtA3Fw} 

\bigskip

\noindent for the latest articles applying the method.  Note also the important caveat that the method trades simplicity against resolution (and often numerical precision), in the sense that the often rather gross approximation of a single-material object will often break down when one seeks to image at sufficiently high resolution, or in many (but not all) imaging scenarios where quantitative information is required.  In such circumstances, more sophisticated approaches---such as the holotomography method reported in Cloetens {\em et al.} (1999)---are required.  Note also that an approximate solution derived {\em e.g.}~using the transport of intensity equation may always be used as starting point for a more sophisticated iterative reconstruction (Gureyev {\em et al.}, 2004).   

\section{Practice}\label{sec:Practice}

\subsection{A quick survey of modern X-ray phase-contrast imaging methods}

Here we describe a few phase contrast methods that have received attention in recent years, and are currently used in the X-ray imaging community. This survey is by no mean exhaustive, and does not cover all the experimental details of the various techniques. A recent key review on the subject, providing insights on the relative strengths and weaknesses of each method, is in Wilkins \textit{et al.} (2014). Bravin \textit{et al.} (2013) not so long ago published a review discussing preclinical and clinical applications of phase contrast imaging. 

Our approach is to look at the different techniques following the Transport of Intensity Equation (TIE), and specifically its version valid for small object-to-detector propagation distances $\delta z$, as written in Eq.~\ref{eq:TIE_for_defocus}. To make our approach clearer, we expand the divergence operator on the right hand side of Eq.~\ref{eq:TIE_for_defocus}:
\begin{eqnarray}\label{eq:TIE_for_defocus2}
I(x,y,z+\delta z) \approx I(x,y,z) \qquad\qquad\qquad\qquad \\ \nonumber -\frac{\delta
z}{k} \left[ \nabla_{\perp}I(x,y,z) \cdot \nabla_{\perp}\phi(x,y,z) + I(x,y,z) \nabla_{\perp}^{2} \phi(x,y,z) \right].
\end{eqnarray}

Under the approximations used to derive this finite-difference version of the TIE, the terms in the square brackets describe the phase contrast contribution to the image. (i) The phase gradient corresponds to the direction of a local stream line (Fig.~\ref{fig:XRayWaveFronts}), whereas (ii) the Laplacian measures the curvature of the wave front (Fig.~\ref{fig:RRR1}). Stated differently: (i) The first term in square brackets contains the (transverse) phase gradient, and represents a prism-like effect that transversely displaces optical energy in a manner proportional to the local deflection angle $\nabla_{\perp}\phi/k$.  (ii) The second term in the square brackets is a lensing term that contains the transverse Laplacian, which describes the local concentration or rarefaction of optical energy density (and hence intensity) due to the sample locally focusing or defocusing the X-ray radiation streaming through it ({\em cf.}~features $J$ and $L$ in Fig. \ref{fig:RRR1}). With the exception of direct methods to measure the phase---such as interferometry---many (but certainly not all!) commonly used phase contrast methods measure phase derivatives, and many such methods can be described using Eq.~\ref{eq:TIE_for_defocus2}.

Methods such as X-ray grating interferometry (Momose \textit{et al.} 2003, Weitkamp \textit{et al.} 2005, Pfeiffer \textit{et al.} 2008) or analyser-based X-ray imaging (Chapman \textit{et al.} 1997, Wernick \textit{et al.} 2003, Rigon \textit{et al.} 2007) provide image contrast dependent upon the first derivative of the phase in the transverse plane (first term in the square bracket). Propagation-based methods (Snigirev \textit{et al.} 1995, Cloetens {\em et al.} 1996, Wilkins \textit{et al.} 1996) measure the second derivative of the phase, described by the second term in the square brackets.

Before going into some more detailed analysis it is worth pointing out two general facts about phase contrast X-ray imaging, which descend straight  from Eq.~\ref{eq:TIE_for_defocus2}.
\begin{description}
\item[Fact 1] All phase contrast imaging techniques require propagation. 
\item [Fact 2] Both gradient and Laplacian of the phase can be present, at the same time, in phase contrast images.
\end{description}

Fact 1 is the obvious consequence of the observation that the second line of Eq.~\ref{eq:TIE_for_defocus2} vanishes if $\delta z=0$. More physically, phase contrast signal---for cases when it is not generated by interferometry---is generated by refraction. X-rays passing through a sample are refracted as well as absorbed. Normally refraction effects goes unnoticed as the refraction angle is extremely small. To become appreciable, measuring refraction requires the detector to be placed some distance away from the sample to analyse the wave front.

Fact 2 is strictly speaking correct only for methods that are sensitive to the phase gradient. Since all of these methods still requires propagation, they will always measure a combination of gradient and Laplacian of the phase (Pavlov {\em et al.} 2004, 2005; Diemoz {\em al.}, 2017). 

\subsection{Phase gradient methods}
In this category we find methods such as analyser-based imaging (ABI), grating interferometry (GI) and its variants, edge illumination (EI) and its variants (Olivo \textit{et al.} 2011, Munro \textit{et al.} 2013), as well as speckle tracking (B\'{e}rujon \textit{et al.} 2012, Morgan \textit{et al.} 2012, Zdora 2018). At the same time, scanning methods using a focused beam as probe (Sayre and Chapman 1995, Schneider 1998) also yield phase gradients and can be included in this description.

Looking once again at Eq.~\ref{eq:TIE_for_defocus2}, we immediately understand what all of these methods have in common. To measure the phase transverse gradient $\nabla_{\perp} \phi(x,y,z)$ one must introduce a transverse intensity gradient $\nabla_{\perp} I(x,y,z)$ and allow for some propagation distance $\delta z$. Interestingly, such an intensity gradient can be introduced in a single frame or throughout multiple frames. Techniques such as speckle tracking or single-image phase retrieval using a grating before the object work by introducing spatial intensity variations in the field of view. Methods such as ABI or grating interferometry (for instance when using diffraction-enhanced imaging or fringe scanning respectively) rely on intensity gradients generated across several images. In this case the phase retrieval method will require more that one image to work. Typically single-image methods are quicker and enable lower X-ray dose, while multi-image methods can attain better spatial resolution.

Another case of multi-image phase retrieval is represented by scanning methods. In this case the intensity gradient is given by the beam itself, that can be shaped either before ({\em i.e.}~in STXM, Scanning Transmission X-ray Microscopy) or after the sample (EI) to yield the desired phase gradient.

Given the common basis we are here describing, it is not surprising that different methods may share similar approaches. In the next subsection we will focus on the complementarity between ABI and GI, showing how similar approaches have been discovered and applied independently by the two communities.

\subsection{Analyser-based and grating-based imaging}

In the spirit of the unified view of gradient-based phase contrast imaging methods, we discuss here in more detail the strong analogy between two popular methods for X-ray phase-contrast imaging, namely ABI and GI. Both methods, in their general form, work by realising an angular scan of either a crystal or one of the gratings. Note that a relative transverse shift $x$ of a grating with respect to the other in GI can be seen as a relative angular change $x/p$ where $p$ is the grating period.

\begin{figure}
\includegraphics[scale=0.5]{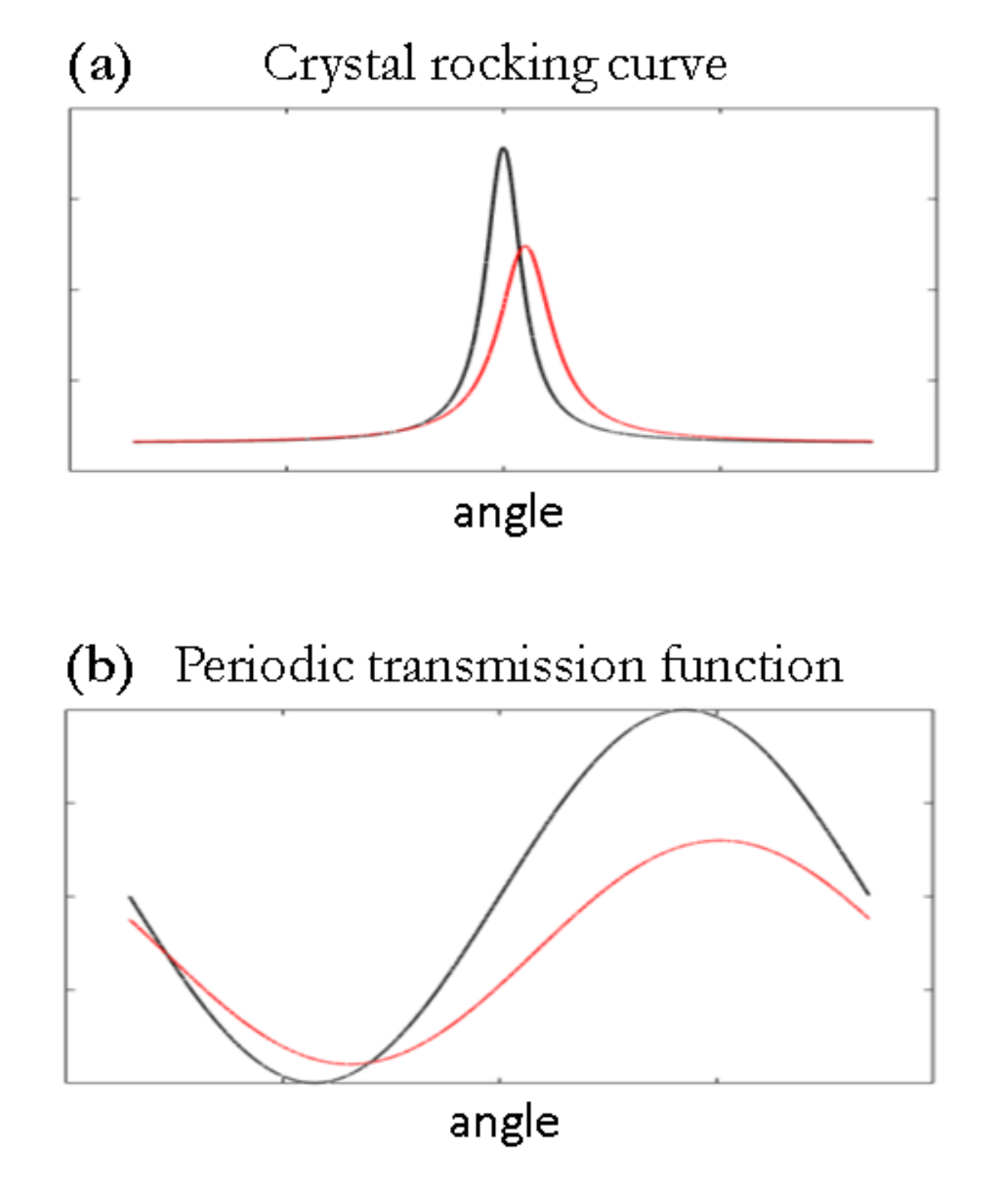}
\caption{(a) Sketch of the transmission function of an ABI system (rocking curve). The black curve represents the rocking curve without the sample, while the red curve represents the rocking curve with the sample. The sample action is to attenuate, refract and scatter the incident beam. (b) Corresponding (simplified) transmission function for a GI system based on two gratings (optionally with a source grating). In this case the transmission function is in general a periodic Moir\'{e} pattern.}
\label{fig:transm_func}
\end{figure}

The analogy therefore begins by considering the angular transmission function of both systems: a rocking curve for ABI and a periodic transmission function for GI (see Fig.~\ref{fig:transm_func}). Historically the development of ABI and GI followed two different approaches, where ABI was brought to fame by a two-image phase retrieval method developed by Chapman \textit{et al.} (1997), while GI phase retrieval was initially based on the fringe scanning (also known as phase stepping) technique (Momose \textit{et al.} 2003, Weitkamp \textit{et al.} 2005). The fringe scanning method is based on the assumption of periodicity of the GI transmission function. A rocking curve used in ABI on the other hand is not periodic, and therefore fringe scanning is unique to GI.

Thus, with the exception of fringe scanning, we can draw a strong analogy between other phase retrieval methods independently developed for ABI and GI. We can divide these methods into two categories, which we will here call the \textit{geometric approach} and \textit{convolution approach} respectively.

The geometric approach is based on Taylor expansion of the transmission function $T$. We here consider the transmission function to be a function of the angle which an X-ray makes with respect to the optic axis; while there are two such angles, corresponding to each of the two independent orthogonal transverse directions, for simplicity we here consider $T$ to be a function of only one angle.  Taylor expansion truncated at the first order means that the transmission function, considered as a function of angle, is locally approximated by a straight line. The intensity transmitted at each point by a sample in this case can be approximated by:
\begin{equation}
I(\theta) \approx I_{0}(\theta) \, T(\theta_0-\Delta \theta_R).
\label{eq:Daniele1}
\end{equation}
Here $ I_{0}(x,y)$ is the intensity before the sample, $\Delta \theta_R$ is the angular shift due to local refraction and $\vert T \vert <1$ accounts for the angle-dependent attenuation. This approximation holds for instance for the flanks of the rocking curve (hence the DEI\footnote{``Diffraction enhanced imaging''.} method by Chapman \textit{et al.} 1997) or the linear part of the sinusoidal transmission function of the GI (hence the two-image methods developed in a tomographic setup for GI by Zhu \textit{et al.} 2010).

A better approximation is represented by truncating the Taylor series at the quadratic term, which means approximating the transmission function locally as a parabola. In this case the sample action is modelled through a combination of attenuation, refraction and scattering, and the intensity after the sample:
\begin{equation}
I(\theta) \approx I_{0}(\theta) \, \int T(\theta_0-\Delta \theta_R -\Delta \theta_S) \, f(\Delta \theta_S) \mathrm{d{\Delta \theta_S}}.
\end{equation}
In this case the sample is assumed not only to attenuate and refract, but also to scatter with a typical scattering width $\Delta \theta_S$. A typical imaging detector is insensitive to such an angular spread generated by scattering, which therefore will be integrated in detection. It is however possible to separate attenuation, refraction and scattering width by acquiring at least three images. A three-image algorithm was  developed by Rigon \textit{et al.}~(2007) in ABI and by Pelliccia {\em et al.}~(2013) in GI. The algorithm is in fact formally the same in both cases.
\\

The methods described above rely on a Taylor expansion of the transmission function. A more accurate approach, which on the other hand requires more images, is represented by the convolution approach. In this case the intensity after the sample is modelled by the convolution product
\begin{equation}
I(\theta) \approx I_{0}(\theta) \star S(\theta),
\end{equation}
where the overall effect of the sample is modelled by the function $S(\theta)$ which has the effect of attenuating, shifting and broadening the transmitted intensity. A phase retrieval algorithm based on the convolution approach was proposed by Wernick \textit{et al.}~(2003) for ABI and by Modregger \textit{et al.}~(2012) in GI. In both cases the algorithm works by determining the shape of $S(\theta)$ by a deconvolution procedure. The technique obviously requires more than three images to produce a reliable estimate of the system transmission function, but has the advantage of increased accuracy (no Taylor expansion involved) in the phase retrieval process.  It is worth noting that the convolution method reduces to the geometric method in the limit specified by Eq.~\ref{eq:Daniele1} (Pelliccia {\em et al.},~2013). 

One can also perform Taylor expansions of the {\em complex transfer function}, to first order (Paganin {\em et al.}, 2004b) and second order (Pavlov {\em et al.}, 2004 and 2005) in spatial frequency, for arbitrary shift invariant linear imaging systems, in the contexts of both the forward problem of generalised phase contrast and the associated inverse problem of phase retrieval.  Thus, for example, a first-order Taylor expansion of the complex transfer function $T(k_x,k_y,\tau_1,\tau_2,\cdots)$ in Eq.~\ref{eq:ArbitraryShiftInvariantLinearCoherentImagingSystem4}, about a Fourier-space point $(k_x^0,k_y^0)$ located near the centre of the Fourier transform of the input wave-field, would approximate the transfer function as $T\approx \gamma_1+\gamma_2(k_x-k_x^0)+\gamma_3(k_y-k_y^0)$, where the constants $\gamma_{1,2,3}$ are functions of the aberration coefficients $\tau_1,\tau_2,\cdots$.  This approach yields equations for the output intensity, obtained from a fairly general perspective, that are very closely related to the ``phase gradient methods'' considered earlier (Paganin {\em et al.}, 2004b).  Similarly, if the transfer function is expanded to second order in spatial frequency, using $T\approx \gamma_1+\gamma_2(k_x-k_x^0)+\gamma_3(k_y-k_y^0)+\gamma_4(k_x-k_x^0)^2+\gamma_5(k_y-k_y^0)^2+\gamma_6(k_x-k_x^0)(k_y-k_y^0)$, one obtains equations for the output intensity that simultaneously exhibit both phase-gradient contrast, and Laplacian-type phase contrast (Pavlov {\em et al.}, 2004 and 2005).

\part{Partial coherence for arbitrary phase contrast imaging systems}

The final part considers partial coherence in the context of arbitrary linear (phase contrast) imaging systems.  

\section{Theory}

\subsection{Partial coherence}

The reader is assumed to be familiar with elementary concepts of partial coherence, including the concepts of spatial and temporal coherence. Hence we assume a more advanced (albeit very intuitive, if one thinks about it for long enough!) perspective based on the coherence-optics equivalent of the density-matrix formalism of quantum mechanics.  This formalism is known as the space--frequency description of partial coherence (Wolf, 1982; see also Mandel and Wolf, 1995, and references therein).

Our intent here is rather modest.  We will briefly introduce the space--frequency description of partial coherence, in which a given partially coherent field may be described via a two-point correlation function known as the {\em cross spectral density}.  The cross-spectral density may be obtained at any given angular frequency via a suitable averaging procedure over {\em an ensemble of strictly monochromatic fields, all of which have the same angular frequency}\footnote {{\em Cf.}~the Gibbs-type statistical ensembles that are often used in the formalism of thermodynamics.}.  We then use this space--frequency description of partially coherent X-ray fields to describe the action of any phase contrast imaging system, provided that the system is linear and the effects of polarisation may be neglected.  

An advantage of the particular formalism adopted here, apart from its broad applicability to a rich variety of both existing and future phase-contrast X-ray imaging and other coherent X-ray imaging scenarios, is the readiness with which the resulting mathematical expressions may be implemented in computer code. Indeed, much of the mathematics that will follow is, in essence, computer pseudo-code rather than explicit calculations {\em per se}.

Since the context is phase-contrast imaging, or imaging more broadly, we describe a given paraxial complex scalar partially coherent X-ray field via a stochastic process that can be characterised via an ensemble of strictly monochromatic fields 
\begin{equation}
\{\psi_\omega^{(j)}(x,y),c_j\}
\label{eq:EnsembleStrictlyMonochromaticFields1}
\end{equation}
at each angular frequency $\omega$, with all fields in each ensemble having the same angular frequency $\omega$ (Wolf, 1982; Mandel and Wolf, 1995).  Here, the {\em j} index labels each member of the ensemble, with the associated real statistical weight $c_j$.  Each of these weights lies between zero and unity inclusive, with 
\begin{equation}
\sum_j c_j = 1.  
\label{eq:EnsembleStrictlyMonochromaticFields2}
\end{equation}
In general, these weights will depend on angular frequency, although for compactness our notation does not explicitly indicate this dependence. 

The {\em cross spectral density} $W_{\omega}(x_1,y_1,x_2,y_2)$ is a two-point correlation function that quantifies the degree of correlation between the optical disturbance at the pair of points $(x_1,y_1)$ and $(x_2,y_2)$, at the specified angular frequency $\omega$.  It is given by the ensemble average over all of the $N$ members of the statistical ensemble:
\begin{eqnarray}
\nonumber W_{\omega}(x_1,y_1,x_2,y_2) &=& \sum_{j=1}^{j=N} c_j \psi_\omega^{(j)*}(x_1,y_1) \psi_\omega^{(j)}(x_2,y_2)  \\ &\equiv& \langle \psi_\omega^{(j)*}(x_1,y_1) \psi_\omega^{(j)}(x_2,y_2)\rangle_{\omega}.
\label{eq:Coherence1}
\end{eqnarray}

The associated spectral density, which may be viewed as the ``diagonal of the cross spectral density'', is:
\begin{equation}
S_{\omega}(x,y) \equiv W_{\omega}(x,y,x,y) = \langle |\psi_\omega^{(j)}(x,y)|^2 \rangle_{\omega}.
\label{eq:GettingSfromW}
\end{equation}

Note that the spectral density is simply the average intensity, being a weighted average over the intensities of the individual fields that make up the statistical ensemble $\{\psi_{\omega}^{(j)}(x,y),c_j\}$.  We note also that the measured intensity $I(x,y)$, registered by a detector which integrates over angular frequencies $\omega$, will be given by the following weighted average of the spectral density:
\begin{equation}\label{eq:Coherence7}
I(x,y)=\int S_{\omega}(x,y) \aleph(\omega) d\omega ,
\end{equation}

\noindent where $\aleph(\omega)$ is a measure of the variable efficiency of the detector, as a function of angular frequency $\omega$.  Note that the shape of the energy spectrum is implicitly taken into account in the cross-spectral density itself, considered as function of energy $E=\hbar\omega$ where $\hbar$ is Planck's constant $h$ divided by $2\pi$.  Note also that the set of weights $\{c_j\}$, used to construct each of the $\omega$-dependent quantities $S_{\omega}(x,y)$ in the above spectral sum, will in general be {\em different} for each $\omega$.  Thus, for the purposes of this paragraph only, $c_j$ would be better notated as $c_{j,\omega}$.  Notwithstanding this, for the sake of notational simplicity we will denote the statistical weights at a given fixed angular frequency $\omega$, as $c_j$, henceforth.  Lastly, we point out that the energy spectrum in the above equation is implicit in the normalisation of the $\psi_{\omega}^{(j)}(x,y)$ functions; thus, if $\omega$ takes a value $\tilde{\omega}$ which is such that there is very little X-ray power at energy $\hbar\tilde{\omega}$, all of the $\psi_{\tilde{\omega}}^{(j)}(x,y)$ functions will be multiplied by a suitably small normalisation factor.

We will take the cross spectral density as the descriptor of partially coherent X-ray fields whose statistics are independent of time (a property known as {\em statistical stationarity}).  We have also implicitly made the assumption of {\em ergodicity}, namely the assumed equality of ensemble averages with time averages. Moreover, if Gaussian statistics may be assumed, then all higher-order correlation functions may be determined from the two-point field correlation function $W$ with which we are working. Proper treatments of these important points will not be given here; they are available in standard texts such as that of Mandel and Wolf (1995).  

\subsection{Modelling a wide class of partially coherent X-ray phase contrast imaging systems}

We now turn to the question of how the space--frequency description for partial coherence may be used to study how cross-spectral densities are {\em transformed upon passage through linear imaging systems utilising partially coherent X-ray radiation}.  Given the previously-mentioned trivial relation between cross-spectral density and spectral density, and the fact that spectral density may be averaged over angular frequency to give the total detected intensity distribution, the formalism outlined below will allow one to {\em determine the intensity distribution output by any linear phase contrast imaging system}.  This covers a rather broad class of existing X-ray phase contrast imaging systems as well as many other coherent-X-ray-optics imaging systems, and we dare say that it also covers a rather large class of coherent imaging systems that will be developed in the future.  While we ignore the effects of polarisation, these can be readily taken into account in a generalisation of the ideas presented here, if required. 

Rather than writing down mathematical expressions, we instead list a sequence of steps that is able to calculate the intensity distribution produced by an arbitrary linear X-ray phase contrast imaging system, illuminating an object, utilising partially coherent radiation under the space--frequency description of partial coherence:  

\begin{itemize}
\item Assume a particular realistic model of the X-ray source after it has passed through whatever conditioning optics, such as monochromators, that may be present in a given X-ray phase contrast imaging system.  This will imply a given statistical ensemble of strictly monochromatic fields
\begin{equation}
\{\psi_\omega^{(j)}(x,y),c_j\}, 
\end{equation}
at the entrance surface of an object that is to be imaged.  

(a) One simple example, of such a statistical ensemble, is an ensemble of monochromatic X-ray plane waves whose directions of propagation are uniformly distributed within some narrow cone whose axis coincides with the optic axis (Pelliccia and Paganin, 2012).  

(b) A second example, of a suitable ensemble of strictly monochromatic fields, could be an ensemble of $z$-directed plane waves that each have a spatially random continuous phase perturbation imprinted upon them (Irvine {\em et al.}, 2010; Morgan {\em et al.}, 2010).  This spatially random phase perturbation could be chosen to have a specified transverse correlation length, and a specified root-mean-square phase excursion.  This provides an excellent model, for example, for partially coherent X-ray imaging systems that utilise a random phase diffuser to ``clean up'' an illuminating beam that contains unwanted structure on account of imperfect optics. 

\item Assume that the object is static\footnote{This ignores potentially important effects such as a moving sample, or the time-dependent accumulation of radiation damage by a radiation-sensitive sample.}, non-magnetic\footnote{This allows polarisation effects to be ignored, for example.}, elastically scattering\footnote{This enables the assumption that the energy of the X-rays is not changed upon passing through the sample.  We thereby ignore often-important phenomena such as fluorescence.} and sufficiently gently spatially varying\footnote{If the projection approximation breaks down, one could employ a better approximation, such as the multi-slice method that was mentioned earlier in these notes.} for the projection approximation to be valid.  The ensemble of fields $\{\psi_\omega^{(j)}(x,y),c_j\}$ at the entrance surface of the object then leads to the ensemble of fields
\begin{equation}
\quad\quad \{\psi_\omega'^{(j)}(x,y),c_j\}=\{\psi_\omega^{(j)}(x,y)\mathcal{T}^{(j)}_{\omega}(x,y),c_j\} 
\end{equation}
at the nominal exit surface of the object, where $\mathcal{T}^{(j)}_{\omega}(x,y)$ is the complex transmission function corresponding to illumination of the object by the strictly monochromatic field $\psi_\omega^{(j)}(x,y)$.  For sufficiently paraxial fields, $\mathcal{T}^{(j)}_{\omega}(x,y)$ may be taken to be independent of $j$, and equal to the complex transmission function that is predicted by the projection approximation, for a $z$-directed plane wave of angular frequency $\omega$.    

\item Each member in the ensemble of fields $\{\psi_\omega'^{(j)}(x,y),c_j\}$ at the exit surface of the object may then be individually propagated through a specified linear imaging system, to give the ensemble of strictly monochromatic fields
\begin{eqnarray}
\{\psi_\omega''^{(j)}(x,y),c_j\} \quad\quad\quad\quad\quad\quad\quad\quad\quad\quad\quad\quad \\ \nonumber =\{\mathcal{D}^{(j)}_{\omega}(\tau_1(\omega),\tau_2(\omega),\cdots)\psi_\omega'^{(j)}(x,y),c_j\}
\end{eqnarray}
at the exit surface of the imaging system. Here,
\begin{eqnarray}
\mathcal{D}^{(j)}_{\omega}(\tau_1(\omega),\tau_2(\omega),\cdots) 
\end{eqnarray}
denotes the operator which maps the {\em j}th input field at angular frequency $\omega$, to the {\em j}th output field at the same angular frequency, when the imaging system is in a state characterised by the control parameters $\tau_1(\omega),\tau_2(\omega),\cdots$.  Note that the control parameters of the imaging system will in general vary with angular frequency, even though the said imaging system would typically have the same physical configuration for all angular frequencies.  However, for paraxial fields, one could often assume $\mathcal{D}^{(j)}_{\omega}(\tau_1(\omega),\tau_2(\omega),\cdots)$ to be independent of $j$.

\item The exit-surface, of the previously mentioned generalised imaging system, is assumed to coincide with the surface of the detector.  The cross-spectral density $W$, the spectral density $S$, and the average intensity $I$, over the surface of the detector, can then be calculated using the ensemble $\{\psi_\omega''^{(j)}(x,y),c_j\}$, together with the previously specified formulae for $W$, $S$ and $I$.     
\end{itemize}

\begin{figure}
\includegraphics[scale=0.15]{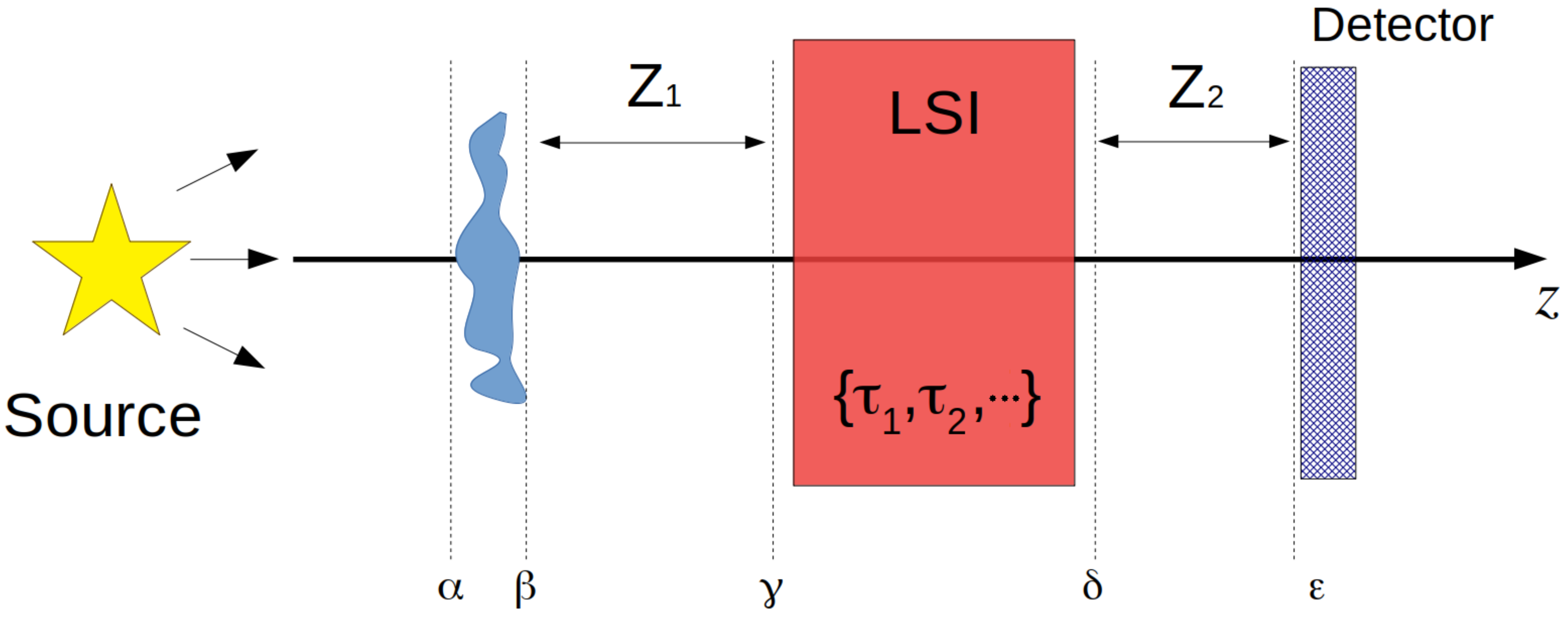}
\caption{Sample experimental setup, to illustrate the means for modelling X-ray imaging systems that is outlined in the main text.}
\label{fig:ExampleImagingSystem}
\end{figure}

As an example of the logic outlined in general terms above, consider the X-ray phase-contrast imaging setup that is sketched in Fig.~\ref{fig:ExampleImagingSystem}.  Here, a monochromated X-ray source (energy $E=\hbar\omega$) illuminates a thin object, before passing through a linear shift-invariant imaging system (LSI), and having the resulting intensity distribution being measured by a position sensitive two-dimensional detector.  

\begin{description}

\item[Example for strictly monochromatic case] 

If one can assume a strictly monochromatic paraxial field and a thin object, the projection approximation can be used to map the complex field $\psi_{\omega}(x,y)$ over the entrance-surface $\alpha$ of the object, to the field 
\begin{equation}
\mathcal{T}_{\omega}(x,y)\psi_{\omega}(x,y)\textrm{ :  plane }\beta 
\end{equation}
over the exit surface $\beta$ (see Eqs~\ref{eq:PhaseShiftProjectionApproximation} and \ref{eq:zz8a} for how to calculate the complex transmission function $\mathcal{T}_{\omega}(x,y)$ under the projection approximation). To propagate through vacuum by the distance $Z_1$ between the exit surface $\beta$ of the object and the entrance surface $\gamma$ of the linear shift-invariant imaging system (LSI system), one can apply the free-space diffraction operator $\mathcal{D}_{Z_1}$ to the field over the plane $\beta$: see Eqs~\ref{eq:FresnelDiffraction} and \ref{eq:DiffractionOperatorFresnelDiffraction}.  This gives:
\begin{equation}
\quad\quad\mathcal{D}_{Z_1}\mathcal{T}_{\omega}(x,y)\psi_{\omega}(x,y)\textrm{ :  plane }\gamma 
\end{equation}
over the entrance surface $\gamma$ of the LSI system.  [As an aside, we re-iterate the fact that operators act from right to left, thus {\em e.g.}~the above equation should be read as meaning ``take the complex field $\psi_{\omega}(x,y)$ over then plane $\alpha$, then multiply by the complex transmission function of the object that is given by $\mathcal{T}_{\omega}(x,y)$ under the projection approximation, then apply the diffraction operator $\mathcal{D}_{Z_1}$ to the resulting field so as to propagate it through a distance of $Z_1$.''] Assuming the LSI system to be in a state characterised by the control parameters $(\tau_1,\tau_2,\cdots)$, one can then use Eq.~\ref{eq:ArbitraryShiftInvariantLinearCoherentImagingSystem5} to propagate to the plane $\delta$, over which the complex disturbance will be:
\begin{equation}
\quad\quad\mathcal{D}(\tau_1,\tau_2,\cdots)\mathcal{D}_{Z_1}\mathcal{T}_{\omega}(x,y)\psi_{\omega}(x,y)\textrm{ :  plane }\delta . 
\end{equation}
Free-space propagation through a distance $Z_2$ then gives the complex disturbance over the surface of the detector:
\begin{equation}
\quad\quad\mathcal{D}_{Z_2}\mathcal{D}(\tau_1,\tau_2,\cdots)\mathcal{D}_{Z_1}\mathcal{T}_{\omega}(x,y)\psi_{\omega}(x,y)\textrm{ :  plane }\epsilon .
\end{equation}
The squared modulus of the above expression gives the intensity measured by the detector.  Alternatively, one could calculate the phase $\phi_{\omega}(x,y,z=\epsilon)$ of the above expression, to determine the phase of the coherent X-ray wave-fronts impinging upon the detector.  Other quantities can also be derived from the above expression for the complex field $\psi_{\omega}(x,y,z=\epsilon)$ over the plane $\epsilon$ coinciding with the surface of the detector, such as the transverse energy-flow vector (Poynting vector)
\begin{equation}
    \quad\quad{\bf S}_{\perp}(x,y) \propto |\psi_{\omega}(x,y,z=\epsilon)|^2 \, \nabla_{\perp}\phi_{\omega}(x,y,z=\epsilon)
\end{equation}
or other related quantities such as the angular-momentum density and the vorticity of the transverse energy-flow vector (see {\em e.g.}~Berry 2009 for how these last-mentioned quantities may be calculated). 

\item[Example for partially-coherent case] The above case showed how to propagate the strictly monochromatic field $\psi_{\omega}(x,y)$ over the entrance surface $\alpha$ of the object in Fig.~\ref{fig:ExampleImagingSystem}, through both the object and an LSI system, to the surface $\epsilon$ of a two-dimensional detector.  If the field is partially coherent, then for each angular frequency $\omega$ corresponding to each energy $E$ via $E=\hbar\omega$, one can instead have an {\em ensemble} of strictly monochromatic fields of the form specified by Eqs~\ref{eq:EnsembleStrictlyMonochromaticFields1} and \ref{eq:EnsembleStrictlyMonochromaticFields2}.  Each field $\psi_{\omega}^{(j)}(x,y)$ is {\em propagated through the optical system in exactly the same way as for the coherent case}, with the associated statistical weights $c_j$ being unchanged via propagation through the optical system.  [Note, however, that while the weights $c_j$ do not change, the normalisation of the individual wave functions may change, as they may be transmitted through the sample or LSI system with different efficiency\footnote{We thank C. Detlefs for bringing this point to our attention.}.] This immediately yields an expression for converting the ensemble of strictly monochromatic input fields (over plane $\alpha$) in Eq.~\ref{eq:EnsembleStrictlyMonochromaticFields1}, to the following ensemble of output fields over the plane $\epsilon$: 
\begin{equation}
\quad\quad\{\mathcal{D}_{Z_2}\mathcal{D}(\tau_1,\tau_2,\cdots)\mathcal{D}_{Z_1}\mathcal{T}_{\omega}(x,y)\psi_{\omega}^{(j)}(x,y),c_j\}.
\label{eq:EnsembleofFieldsOverDetector}
\end{equation}
The resulting statistical ensemble can be used to calculate a number of derived quantities of interest:

\begin{itemize}
\item The cross-spectral density $W_\omega(x_1,y_1,x_2,y_2)$ (see Eq.~\ref{eq:Coherence1});
\item The spectral density $S_\omega(x,y)$ (see Eq.~\ref{eq:GettingSfromW}), which may be integrated over $\omega$ to give the total spectral density (see Eq.~\ref{eq:Coherence7});
\item The position-dependent ensemble-averaged angular momentum density and vorticity, via ensemble averaging the expressions for these quantities that appear in a coherent-optics context in Berry (2009);
\item The Wigner function, the ambiguity function, the generalised radiance function {\em etc.}~are two-point correlation functions that can all be derived from the ensemble in Eq.~\ref{eq:EnsembleofFieldsOverDetector}\footnote{See {\em e.g.}~Alonso (2011) for a description of how the Wigner, ambiguity, generalised-radiance and related functions may be calculated from the statistical ensemble of fields in Eq.~\ref{eq:EnsembleofFieldsOverDetector}.}. This amounts to a computationally simple means for modelling the important ``coherence transport'' problem, especially when the just-mentioned correlation functions are calculated for different points along the optic axis, as one traverses the various optical elements in sequence from the source to the detector. 

\end{itemize}
\end{description}

The above formalism may also be trivially generalised to the case of four-point correlation functions, six-point correlations functions and so on.  Four-point correlation functions are primarily of interest in the X-ray context of the Hanbury Brown--Twiss effect, which plays a role in X-ray ghost imaging (Yu {\em et al.}~2016; Pelliccia {\em et al.}~2016; Schori and Shwartz 2017; Zhang {\em et al.}~2018; Pelliccia {\em et al.}~2018; Schori {\em et al.}~2018; Ceddia and Paganin 2018; Gureyev {\em et al.}~2018; Kingston {\em et al.}~2018; Kim {\em et al.}~2018; Kingston {\em et al.}~2019).  While interesting, X-ray ghost imaging will not be discussed in these tutorials.  Also, as previously mentioned, for the common special case of Gaussian statistics, all higher-order correlation functions are either determined by the two-point correlation functions that have been our main focus, or they vanish (see {\em e.g.}~Mandel and Wolf, 1995).  We also note that the correlation functions of various orders would be of pivotal importance in the context of the nascent field of quantum X-ray optics (Adams {\em et al.}, 2013; Kuznetsova and Kocharovskaya, 2017), which has yet to mature to the same stage of development as quantum optics in the context of visible light.  This maturation of quantum X-ray optics in coming years is an obvious emerging research opportunity for coming generations, as one can readily ascertain by studying the most recent literature.
 
\subsection{Speculations regarding future trends} 
Below we give a sequence of speculations regarding possible future trends in X-ray phase contrast imaging in particular, and coherent X-ray imaging more broadly.  These speculations, which range from the self-evident to the tentatively hypothetical, are intended as both stimulus and guide for future research in the field. 

(a) As one moves to progressively higher resolution, {\em e.g.} in tomography, the projection approximation will become increasingly ill behaved.  One will eventually need to embrace fully dynamical models such as the multi-slice approximation, in the context of the inverse problem of tomography, or diffraction-tomography methods, to a larger degree than is the case at present.  This transition is already in progress, as even a cursory survey of recent literature will show.  On a related note, and again as one moves to ever-higher resolution, the scalar approximation for the X-ray wave-field may begin to break down {\em e.g.} when large scattering angles or magnetic phenomena are considered (Detlefs, 2019).  In all of this, much guidance is to be gleaned from the existing literature on electron tomography, which has---very broadly speaking---been forced to grapple with such problems at an earlier stage than the X-ray tomography community.  Much can also be learned from the very well established field of X-ray scattering and absorption by magnetic materials---see {\em e.g.}~the text by Lovesey and Collins (1996)---together with any aspects of X-ray physics in which the effects of magnetism and/or polarisation are important.

(b) Some emphasis has been given in these notes to X-ray phase contrast imaging using shift-invariant linear imaging systems that possess arbitrary aberrations. This was done, in part, because a large fraction of the existing suite of phase contrast imaging systems ({\em e.g.}~propagation-based phase contrast, Zernike phase contrast, analyser-crystal phase contrast, edge-illumination phase contrast {\em etc.}) may be viewed as aberrated imaging systems, in the sense of possessing a complex transfer function that is not unity.  This emphasis also had the following future speculation in mind: As coherent X-ray imaging systems become more mature, the ability to tune particular aberrations will become increasingly refined, in light of the fact that non-zero aberrations are a necessary condition for the existence of phase contrast.  Rather than considering oneself to be limited to the particular aberrations associated with a particular phase-contrast imaging system, it may be fruitful to instead consider a near future where a suite of different aberration parameters may be accessed or tuned at will.  Again, this speculated future state of affairs---for coherent X-ray imaging---is a current reality for electron imaging, now that transmission electron microscopes that are corrected for spherical aberration and have tunable aberrations have become available (Pennycook, 2017).

(c) Here we remark on possible future developments regarding the complementarity between iterative and deterministic methods for solving inverse imaging problems in coherent X-ray optics. 

\begin{itemize}

\item Iterative methods have the virtue of applicability to a wider class of inverse imaging problems, and in particular to imaging problems for which deterministic closed-form solutions are not known.  Iterative methods typically rely on minimisation of an error metric, which is often rather easy to write down but typically rather computationally expensive to minimise.  Such computational expense, which typically seeks error-metric minimisation in a function space of high dimension, will become less of a limitation in light of the availability of cheap ever-more-powerful computing machines.  The ideas of compressive sensing and machine learning are both readily incorporated into iterative approaches, which again speaks in their favour regarding future applications.  One negative of iterative approaches is that they often leave obscure the question of the uniqueness, and therefore the correctness, of the resulting reconstruction.  They also often leave obscure the fundamental questions of the stability of reconstructions with respect to imperfections in the input data.  These negatives must be balanced against the very rich variety of successful iterative image reconstructions in the context of coherent X-ray imaging that are extant, a state of affairs that will surely continue in future for a wide variety of imaging scenarios.  

\item Deterministic approaches seek closed-form solutions to the inverse problems of coherent X-ray optics.  While applicable to a smaller class of (simpler) problems compared to the larger set of problems that may be treated iteratively, deterministic approaches often but not always have the advantage of greater conceptual clarity and speed of reconstruction. The conceptual clarity is provided by knowledge that the provided solution is both unique and stable with respect to perturbations in the input data that will be present for any realistic experiment, while speed is provided by application of an often-simple closed-form formula.  It is possible that, in the future, some problems that were hitherto only thought to be soluble via iterative means, may turn out to be addressable by deterministic means, after all.  

\item Lastly, we remark that iterative and deterministic methods are not mutually exclusive, as there are many situations in which the approximate solutions provided by the latter may be iteratively refined by the former.  Both iterative and deterministic methods are important, and the just-described complementarity is likely to prevail in future. See {\em e.g.}~Pavlov {\em et al.}~ (2018) for more information.     

\end{itemize}

(d) The key ideas of compressed sensing, machine learning and artificial intelligence are likely to play an increasing role in both the quantitative and qualitative analysis of X-ray imaging data, on a variety of levels.  For example: 
\begin{itemize}

\item Compressed sensing comes with the promise of significant improvements in the efficiency of X-ray data collection, permitting reconstructions of a sample of interest to be performed, under certain circumstances, with fewer data (and therefore less dose) than was previously thought possible.

\item Machine learning (an approach that is surging massively across many fields) comes with the promise of very holistic and flexible analyses which make fuller use of the totality of available X-ray data, than has been the case hitherto.  In this context we note the illuminating comment of Carsten Detlefs (2019): ``Machine learning depends critically on the availability of training data. In many cases a forward simulation of a known (virtual) sample is much easier to compute than the reconstruction of an unknown sample. Thus, virtual experiments might be used to train machine learning algorithms without the need for (costly) synchrotron experiments.  On the other hand, the quality of the machine analysis depends critically on the quality of the training data. Therefore any bias, systematic error, or failure to represent a systematic error in the real experiment may lead to flaws in the analysis.'' 

\item Artificial intelligence, besides the role it can play in machine learning, brings the promise of automated or semi-automated assessment of large volumes of X-ray imaging data, {\em e.g.}~in a medical-diagnosis or an industrial-testing context.    

\end{itemize}

(e) We have nowhere mentioned speckles, in our discussions relating to partial coherence.  This is remiss, given that the concept of unresolved speckle lies at the heart of most phenomena relating to partial coherence (see {\em e.g.}~Nugent {\em et al.}~2003; Vartanyants \& Robinson 2003;  Nesterets 2008).  The following comment from Vartanyants \& Robinson (2003), regarding unresolved speckle, is particularly illuminating: ``It is important to note that the `decoherence' effect of optics is not a degradation of the inherent source coherence, but instead the creation of an entirely new component to the coherence function with a dramatically reduced coherence length.'' Typical scalar partially coherent X-ray fields are littered with a profusion of speckles that are too rapidly varying in both space and time to be resolved by one's detectors.  The idea that one averages over such speckles, with a spatial average over the detector element and a temporal average over the detection time, gives a means for visualising and indeed modelling partially coherent X-ray fields.  With current computing power, and increased computing power projected for the future, such direct spatio-temporal modelling of partially coherent fields in terms of underpinning fluctuating speckle fields, forms a very promising and conceptually illuminating avenue for future research.  Note that, at the scalar-field level, such speckle fields are threaded by typically complicated---indeed fractal (O'Holleren {\em et al.}, 2008)---random networks of nodal-line-threaded phase singularities associated with X-ray vortices.  For more on X-ray phase vortices and their associated nodal lines (``threads of darkness''), see {\em e.g.}~Chapter 5 of Paganin (2006), together with references therein.  If one ascends to a vectorial description of the classical X-ray electromagnetic field for partially-coherent X-ray imaging scenarios, the speckles in the instantaneous energy density will remain, but the phase vortices will be threaded by a random network of vectorial singularities (see {\em e.g.}~Nye (1999), and references therein, for a discussion of the singularities of vector fields such as the electric and magnetic fields). 

(f) Virtual optics, in which the computer forms an intrinsic component of an X-ray optical imaging system, has of course long been a reality, {\em e.g.}~in X-ray computed tomography (Diemoz {\em et al.}~2017), interferometry (Bonse \& Hart, 1965), holography (Gabor, 1948), crystallography (Hammond, 2001), coherent diffractive imaging (Miao {\em et al.}~1999), ptychography (Pfeiffer, 1999) {\em etc}.  In virtual-optics systems, the computer forms an intrinsic component of the imaging system, with (i) hardware optics streaming and manipulating X-ray optical information at the field level, supplemented by (ii) digital computers that stream and manipulate digital optical information at the level of bits. Stated differently, an analogue computer (optical hardware streaming and manipulating physical X-ray wave-fields) is coupled with a digital computer (digital information processing unit).  As computers become both more powerful and cheaper, it is likely that further impetus will be given to enhancing the role of computers as an intrinsic component of X-ray imaging systems, with a corresponding simplification of the associated optical hardware.  For an example of virtual optics related to X-ray phase contrast and phase retrieval, see {\em e.g.}~Paganin {\em et al.}~(2004a).    

(g) Much thought has been given to the implications of future advances in the brightest-available X-ray source technology, such as X-ray free-electron lasers.  We will not add to this discussion here.  Rather, we explore a different class of implications of advances in source technology, regarding off-the-shelf X-ray sources available to small research facilities, hospitals and industry.  In this context, we state an obvious fact: while third-generation X-ray synchrotron sources will become increasingly divergent from fourth-generation and higher-generation X-ray sources, they are likely to become increasingly convergent towards off-the-shelf compact sources available in smaller scale research laboratories and hospitals.  This is important in contexts such as medical and industrial imaging, enhancing the ability  of proof-of-concept third-generation-synchrotron experiments to be translated to medical and industrial settings.  

(h) What will be the implications of advances in detector technology, for example due to future pixel detectors being able to measure a full X-ray spectrum at each pixel of a two-dimensional image? Surely there are immense future opportunities here, to mine the very information-rich data-sets that such detectors will provide.  

\section*{Acknowledgements}

This tutorial was presented as three two-hour seminars, delivered to the European Synchrotron (ESRF) community, on May 31 -- June 2, 2017. We thank the European Synchrotron for facilitating and videotaping these lectures, as well as making them available to all on YouTube (see link in abstract).  

Both authors offer sincere thanks to Alexander Rack and the ESRF Directors for supporting their visits to ESRF, and for the privilege of working with Alexander Rack, Margie Olbinado and Yin Cheng on X-ray ghost imaging.  

Thanks to all of you who stopped by, for inspiring conversations and meetings during our ESRF visits in 2017 and 2018. 

Figure  \ref{fig:cell} was artistically drawn and coloured by Kristina Pelliccia.

Thankyou to Carsten Detlefs (ESRF), for his close reading of the first arXiv version of this text, and for his detailed comments.  Incorporating these corrections and clarifications has led to many significant improvements.

\section*{References}

B.W. Adams, C. Buth, S.M. Cavaletto, J. Evers, Z. Harman, C.H. Keitel, A. P\'{a}lffy, A. Pic\'{o}n, R. R\"{o}hlsberger, Y. Rostovtsev and K. Tamasaku, {\em X-ray quantum optics}, J. Mod. Opt. {\bf 60}, 2--21 (2013).

M.A. Alonso, {\em Wigner functions in optics: describing beams as ray bundles and pulses as particle ensembles}, Adv. Opt. Photonics {\bf 3}, 272--365 (2011).

M.A. Beltran, D.M. Paganin, K. Uesugi and M.J. Kitchen, {\em 2D and 3D X-ray phase retrieval of multi-material objects using a single defocus distance}, Opt. Express {\bf 18}, 6423--6436 (2010).

M.A. Beltran, D.M. Paganin, K.K.W. Siu, A. Fouras, S.B. Hooper, D.H. Reser and M.J. Kitchen, {\em Interface-specific X-ray phase retrieval tomography of complex biological organs}, Phys. Med. Biol. {\bf 56}, 7353--7369 (2011).

M.V. Berry and C. Upstill, {\em Catastrophe Optics: Morphologies of caustics and their diffraction patterns}, Prog.~Opt. {\bf 18}, 257--346 (1980). 

M.V. Berry, {\em Much ado about nothing: Optical dislocation lines (phase singularities, zeros, vortices ...)}, Proc.~SPIE {\bf 3487}, 1--5 (1998).

M.V. Berry, {\em Optical currents}, J. Opt. A: Pure Appl. Opt. {\bf 11}, 094001 (2009).

S. B\'{e}rujon, E. Ziegler, R. Cerbino and L. Peverini, {\em Two-dimensional X-ray beam phase sensing}, Phys. Rev. Lett. {\bf 108}, 158102 (2012). 

A. Bravin, P. Coan and P. Suortti, {\em X-ray phase-contrast imaging: from pre-clinical applications towards clinics}, Phys. Med. Biol. {\bf58}, R1--R35 (2013).

H. Bremmer, {\em On the asymptotic evaluation of diffraction
integrals with a special view to the theory of defocusing and
optical contrast}, Physica {\bf 18}, 469--485 (1952). 

U. Bonse and M. Hart, {\em A X-ray interferometer}, Appl. Phys. Lett. {\bf 6}, 155--156 (1965).  

D. Ceddia and D.M. Paganin, {\em Random-matrix bases, ghost imaging and x-ray phase contrast computational ghost imaging}, Phys. Rev. A {\bf 97}, 062119 (2018). 

D. Chapman, W. Thomlinson, R. E. Johnston, D. Washburn, E. Pisano, N. Gm\"ur, Z. Zhong, R. H. Menk, F. Arfelli and D. Sayers, {\em Diffraction enhanced X-ray imaging}, Phys. Med. Biol. {\bf 42}, 2015--2025 (1997).

P. Cloetens, R. Barrett, J. Baruchel, J.-P. Guigay and M. Schlenker, {\em Phase objects in synchrotron radiation hard X-ray imaging}, J. Phys. D: Appl. Phys. {\bf 29}, 133--146 (1996). 

P. Cloetens, W. Ludwig, J. Baruchel, D. Van Dyck, J. Van Landuyt, J. P. Guigay and M. Schlenker, {\em Holotomography: Quantitative phase tomography with micrometer resolution using hard synchrotron radiation x rays}, Appl. Phys. Lett. {\bf 75}, 2912--2914 (1999).

J.M. Cowley and A.F. Moodie, {\em The scattering of electrons by atoms and crystals. I. A new theoretical approach}, Acta Cryst. {\bf 10}, 609--619 (1957). 

J.M. Cowley and A.F. Moodie, {\em The scattering of electrons by atoms and crystals. III. Single-crystal diffraction patterns}, Acta Cryst. {\bf 12}, 360--367 (1959). 

C. Detlefs, Private communication to D.M. Paganin, 2019.

F. D\"{o}ring, A.L. Robisch, C. Eberl, M. Osterhoff, A. Ruhlandt, T. Liese, F. Schlenkrich, S. Hoffmann, M. Bartels, T. Salditt and H.U. Krebs, {\em Sub-5 nm hard X-ray point focusing by a combined Kirkpatrick-Baez mirror and multilayer zone plate}, Opt. Express {\bf 21}, 19311--19323 (2013).

P. C. Diemoz, C. K. Hagen, M. Endrizzi, M. Minuti, R. Bellazzini, L. Urbani, P. De Coppi and A. Olivo, {\em Single-shot X-ray phase-contrast computed tomography with nonmicrofocal laboratory sources},  Phys. Rev. Appl. {\bf 7}, 044029 (2017). 

E. F\"{o}rster, K. Goetz and P. Zaumseil, {\em Double crystal diffractometry for the characterization of targets for laser fusion experiments}, Kristall und Technik, {\bf 15} 937--945 (1980). 

D. Gabor, {\em A new microscopic principle}, Nature {\bf 161}, 777--778 (1948). 

G. Gbur and E. Wolf, {\em Relation between computed tomography and diffraction tomography}, J. Opt. Soc. Am. A {\bf 18}, 2132--2137 (2001). 

D. Giovannini, J. Romero, V. Poto\u{c}ek, G. Ferenczi, F. Speirits, S.M. Barnett, D. Faccio and M.J. Padgett, {\em Spatially structured photons that travel in free space slower than the speed of light}, Science {\bf 347}, 857--860 (2015).

J.W. Goodman, {\em Introduction to Fourier Optics}, 3rd edn, Roberts \& Company, Englewood Colorado (2005).

T.E. Gureyev, A. Pogany, D.M. Paganin and S.W. Wilkins, {\em Linear algorithms for phase retrieval in the Fresnel region}, Opt. Commun. {\bf 231}, 53--70 (2004).

T.E. Gureyev, S.C. Mayo, D.E. Myers, Ya.Nesterets, D.M. Paganin, A. Pogany, A.W. Stevenson and S.W. Wilkins, {\em Refracting R\"{o}ntgen's rays: Propagation-based x-ray phase contrast for biomedical imaging}, J. Appl. Phys. {\bf 105}, 102005 (2009).

T.E. Gureyev, S.C. Mayo, Ya.I. Nesterets, S. Mohammadi, D. Lockie, R.H. Menk, F. Arfelli, K.M. Pavlov, M.J. Kitchen, F. Zanconati, C. Dullin and G. Tromba, {\em Investigation of the imaging quality of synchrotron-based phase-contrast mammographic tomography}, J. Phys. D: Appl. Phys. {\bf 47}, 365401 (2014).

T.E. Gureyev, D.M. Paganin, A. Kozlov, Ya.I. Nesterets and H.M. Quiney, {\em Complementary aspects of spatial resolution and signal-to-noise ratio in computational imaging}, Phys. Rev. A {\bf 97}, 053819 (2018).

J. Hadamard, {\em Lectures on Cauchy's Problem in Linear Partial Differential Equations}, Yale University Press, New Haven (1923). 

C. Hammond, {\em The Basics of Crystallography and Diffraction}, 2nd ed.,
Oxford University Press, Oxford (2001).

S.C. Irvine, K.S. Morgan, Y. Suzuki, K. Uesugi, A. Takeuchi, D.M. Paganin and K.K.W. Siu, {\em Assessment of the use of a diffuser in propagation-based phase contrast imaging}, Opt. Express {\bf 18}, 13478--13491 (2010).

Y.Y. Kim, L. Gelisio, G. Mercurio, S. Dziarzhytski, M. Beye, L. Bocklage, A. Classen, C. David, O.Yu. Gorobtsov, R. Khubbutdinov, S. Lazarev, N. Mukharamova, Yu.N. Obukhov, B. Roesner, K. Schlage, I.A. Zaluzhnyy, G. Brenner, R. Roehlsberger, J. von Zanthier, W. Wurth, and I.A. Vartanyants, {\em Ghost Imaging at an XUV Free-Electron Laser}, pre-print available at \url{https://arxiv.org/abs/1811.06855} (2018).

A.M. Kingston, D. Pelliccia, A. Rack, M.P. Olbinado, Y. Cheng, G.R. Myers and D.M. Paganin, {\em Ghost tomography}, Optica {\bf 5}, 1516--1520 (2018).  

A.M. Kingston, G.R. Myers, D. Pelliccia, I.D. Svalbe and D.M. Paganin, {\em X-ray ghost-tomography: Artefacts, dose distribution, and mask considerations}, IEEE Trans. Comput. Imaging {\bf 5}, 136--149 (2019). 

E.J. Kirkland, {\em Advanced Computing in Electron Microscopy}, 2nd ed., Springer, Berlin/Heidelberg (2010).

M.J. Kitchen, G.A. Buckley, T.E. Gureyev, M.J. Wallace, N. Andres-Thio, K. Uesugi, N. Yagi and S.B. Hooper, {\em CT dose reduction factors in the thousands using X-ray phase contrast}, Sci. Rep. {\bf 7}, 15953 (2017).

H.-O. Kreiss and J. Lorenz, {\em Initial-Boundary Problems and the Navier--Stokes Equations}, Academic Press, San Diego (1989).

R. Kress, {\em Linear Integral Equations}, Springer--Verlag, Berlin, p.~221 (1989).

E. Kuznetsova and O. Kocharovskaya, {\em Quantum optics with X-rays}, Nat. Phot. {\bf 11}, 685--686 (2017).

K. Li, M. Wojcik, and C. Jacobsen, {\em Multislice does it all -- calculating the performance of nanofocusing X-ray optics}, Opt. Express {\bf 25}, 1831--1846 (2017).

S.W. Lovesey and S.P. Collins, {\em X-ray Scattering and Absorption by Magnetic Materials}, Oxford University Press, Oxford (1996). 

E. Madelung, {\em Quantentheorie in hydrodynamischer Form}, Z. Phys. {\bf 40}, 322--326 (1927). 

L. Mandel and E. Wolf, {\em Optical Coherence and Quantum Optics}, Cambridge University Press, Cambridge (1995).  

H.E. Martz Jr., B.J. Kozioziemski, S.K. Lehman, S. Hau-Riege, D.J. Schneberk and A. Barty, { \em Validation of radiographic simulation codes including X-ray phase effects for millimeter-size objects with micrometer structures}, J. Opt. Soc. Am. A {\bf 24}, 169--178 (2007).

J. Miao, P. Charalambous, J. Kirz and D. Sayre, {\em Extending the methodology of
X-ray crystallography to allow imaging of micrometre-sized non-crystalline specimens}, Nature {\bf 400}, 342--344 (1999).

P. Modregger, F. Scattarella, B. R. Pinzer, C. David, R. Bellotti and M. Stampanoni, {\em Imaging the ultrasmall-angle X-ray scattering distribution with grating interferometry}, Phys. Rev. Lett. {\bf 108}, 048101 (2012).

A. Momose, S. Kawamoto, I. Koyama, Y. Hamaishi, K. Takai and Y. Suzuki, {\em Demonstration of X-ray Talbot interferometry}, Jpn. J. Appl. Phys. {\bf 42}, L866--L868 (2003).

K.S. Morgan, S.C. Irvine, Y. Suzuki, K. Uesugi, A. Takeuchi, D.M. Paganin and K.K.W. Siu, {\em Measurement of hard x-ray coherence in the presence of a rotating random-phase-screen diffuser}, Opt. Commun. {\bf 283}, 216--225 (2010). 

K.S. Morgan, D.M. Paganin and K.K.W. Siu, {\em X-ray phase imaging with a paper analyzer}, Appl. Phys. Lett. {\bf 100}, 124102 (2012). 

P.R.T. Munro, L. Rigon, K. Ignatyev, F.C.M. Lopez, D. Dreossi, R.D. Speller and A. Olivo, {\em A quantitative, non-interferometric X-ray phase contrast imaging technique}, Opt. Express {\bf 21}, 647--661 (2013).

P. M\"uller, M. Sch\"urmann and J. Guck, {\em The theory of diffraction tomography},  \url{https://arxiv.org/abs/1507.00466} (2016).

Ya.I. Nesterets,  {\em On the origins of decoherence and extinction contrast in phase-contrast imaging}, Opt. Commun. {\bf 281},
533--542 (2008).

Ya.I. Nesterets and T.E. Gureyev, {\em Noise propagation in x-ray phase-contrast imaging and computed tomography}, J. Phys. D: Appl. Phys. {\bf 47} 105402 (2014).

K.A. Nugent, T.E. Gureyev, D. Cookson, D. Paganin and Z. Barnea, {\em Quantitative phase imaging using hard X-rays}, Phys. Rev. Lett. {\bf 77}, 2961--2964 (1996). 

K. A. Nugent, C. Q. Tran and A. Roberts, {\em Coherence transport through imperfect x-ray optical systems}, Opt. Express {\bf 11}, 2323-2328 (2003).

J.F. Nye, {\em Natural Focusing and Fine Structure of Light}, Institute of Physics Publishing, Bristol (1999).

K. O'Holleran, M.R. Dennis, F. Flossmann and M.J. Padgett, {\em Fractality of Light's Darkness}, Phys. Rev. Lett. {\bf 100}, 053902 (2008).

A. Olivo, K. Ignatyev, P.R.T. Munro and R.D. Speller, {\em Noninterferometric phase-contrast images obtained with incoherent X-ray sources}, Appl. Opt. {\bf 50}, 1765--1769 (2011).

D.M. Paganin, S.C. Mayo, T.E. Gureyev, P.R. Miller and S.W. Wilkins, {\em Simultaneous phase and amplitude extraction from a single defocused image of a homogeneous object}, J. Microsc. {\bf 206}, 33--40 (2002).  

D. Paganin, T.E. Gureyev, S.C. Mayo, A.W. Stevenson, Ya.I. Nesterets and S.W. Wilkins, { \em X-ray omni microscopy}, J. Microsc. {\bf 214}, 315--327 (2004a).

D. Paganin, T.E. Gureyev, K.M. Pavlov, R.A. Lewis and M. Kitchen, {\em Phase retrieval using coherent imaging systems with linear transfer functions}, Opt. Commun. {\bf 234}, 87--105 (2004b).

D. Paganin, A. Barty, P.J. McMahon and K.A. Nugent, {\em Quantitative phase-amplitude microscopy. III. The effects of noise}, J. Microsc. {\bf 214}, 51--61 (2004c).  

D.M. Paganin, {\em Coherent X-ray Optics}, Oxford University Press, Oxford (2006).

D.M. Paganin and T.E. Gureyev, {\em Phase contrast, phase retrieval and aberration balancing in shift-invariant linear imaging systems}, Opt. Commun. {\bf 281}, 965--981 (2008).

D.M. Paganin, T.C. Petersen and M.A. Beltran, {\em Propagation of fully coherent and partially coherent complex scalar fields in aberration space}, Phys. Rev. A {\bf 97}, 023835 (2018).

K.M. Pavlov, T.E. Gureyev, D. Paganin, Ya.I. Nesterets, M.J. Morgan and R.A. Lewis, {\em Linear systems with slowly varying transfer functions and their application to x-ray phase-contrast imaging}, J. Phys. D: Appl. Phys. {\bf 37}, 2746--2750 (2004).

K.M. Pavlov, T.E. Gureyev, D. Paganin, Ya.I. Nesterets, M.J. Kitchen, K.K.W. Siu, J.E. Gillam, K. Uesugi, N. Yagi, M.J. Morgan and R.A. Lewis, {\em Unification of analyser-based and propagation-based X-ray phase-contrast imaging}, Nucl. Inst. Meths Phys. Res. A {\bf 548}, 163--168 (2005).

K.M. Pavlov, K.S. Morgan, V.I. Punegov and D.M. Paganin, {\em Deterministic X-ray Bragg coherent diffraction imaging as a seed for subsequent iterative reconstruction}, J. Phys. Commun. {\bf 2}, 085027 (2018).

D. Pelliccia and D.M. Paganin, {\em Coherence vortices in vortex-free partially coherent x-ray fields}, Phys. Rev. A {\bf 86}, 015802 (2012). 

D. Pelliccia, L. Rigon, F. Arfelli, R.H. Menk, I. Bukreeva and A. Cedola, {\em A three-image algorithm for hard X-ray grating interferometry},  Opt. Express, {\bf 21}, 19401--19411 (2013).

D. Pelliccia, A. Rack, M. Scheel, V. Cantelli and D.M. Paganin, {\em Experimental X-ray ghost imaging}, Phys. Rev. Lett. {\bf 117}, 113902 (2016).  

D. Pelliccia, M. P. Olbinado, A. Rack, A. M. Kingston, G. R. Myers and D. M. Paganin, {\em Towards a practical implementation of X-ray ghost imaging with synchrotron light}, IuCrJ {\bf 5}, 428--438 (2018).

S.J. Pennycook, {\em The impact of STEM aberration correction on materials science}, Ultramicroscopy {\bf 180}, 22--33 (2017).

F. Pfeiffer, M. Bech, O. Bunk, P. Kraft, E. F. Eikenberry, Ch. Br\"onnimann, C. Gr\"unzweig and C. David, {\em Hard-X-ray dark-field imaging using a grating interferometer}, Nat. Mater. {\bf 7}, 134--137 (2008)

F. Pfeiffer, {\em X-ray ptychography}, Nat. Photonics {\bf 12}, 9--17 (2018).

L. Rigon, F. Arfelli and R.-H. Menk, {\em Three-image diffraction enhanced imaging algorithm to extract absorption, refraction and ultrasmall-angle scattering}, Appl. Phys. Lett. {\bf 90}, 114102 (2007).

D. Ruelle, {\em Chaotic Evolution and Strange Attractors}, Cambridge University Press, Cambridge (1989).

B.E.A. Saleh and M.C. Teich, {\em Fundamentals of Photonics}, 2nd edn, Wiley-Interscience, Hoboken New Jersey (2007).

D. Sayre and H. N. Chapman, {\em X-ray microscopy}, Acta Cryst. {\bf A51} 237--252 (1995).

P. Schiske, {\em Image reconstruction by means of focus series}, J. Microsc. {\bf 207}, 154 (2002).  Note that this is a translation from the German, of a paper that first appeared in 1968, in the {\em Proceedings of the Fourth Regional Congress on Electron Microscopy},
Rome 1968, vol. 1, pp. 145--146.

G. Schneider, {\em Cryo X-ray microscopy with high spatial resolution in amplitude and phase contrast}, Ultramicroscopy {\bf 75} 85--104 (1998).

A. Schori and S. Shwartz, {\em X-ray ghost imaging with a laboratory source}, Opt. Express {\bf 25}, 14822--14828 (2017). 

A. Schori, D. Borodin, K. Tamasaku and S. Shwartz, {\em Ghost imaging with paired x-ray photons}, Phys. Rev. A {\bf 97}, 063804 (2018).

A. Snigirev, I. Snigireva, V. Kohn, S. Kuznetsov and I. Schelokov, {\em On the possibilities of X-ray phase contrast microimaging by coherent high-energy synchrotron radiation}, Rev. Sci. Instrum. {\bf 66}, 5486--5492 (1995). 

M.R. Teague, {\em Deterministic phase retrieval: a Green's
function solution}, J. Opt. Soc. Am. {\bf 73}, 1434--1441 (1983). 

I.A. Vartanyants and I.K. Robinson, {\em Origins of decoherence in coherent X-ray
diffraction experiments}, Opt. Commun. {\bf 222}, 29--50 (2003).

T. Weitkamp, A. Diaz, C. David, F. Pfeiffer, M. Stampanoni, P. Cloetens and E. Ziegler, {\em X-ray phase imaging with a grating interferometer}, Opt. Express {\bf 13}, 6296--6304 (2005).

M. N. Wernick, O. Wirjadi, D. Chapman, Z. Zhong, N. P. Galatsanos, Y. Yang, J. G. Brankov, O. Oltulu, M. A. Anastasio and C. Muehleman, {\em Multiple-image radiography}, Phys. Med. Biol. {\bf 48}, 3875--3895 (2003).

V. White and D. Cerrina, {\em Metal-less X-ray phase-shift masks for nanolithography}, J. Vac. Sci. Technol. B {\bf 10}, 3141--3144 (1992). 

S.W. Wilkins, T.E. Gureyev, D. Gao, A. Pogany and A.W. Stevenson, {\em Phase-contrast imaging using polychromatic hard X-rays}, Nature {\bf 384}, 335--338 (1996). 

S.W. Wilkins, Ya.I. Nesterets, T.E. Gureyev, S.C. Mayo, A. Pogany and A.W. Stevenson, {\em On the evolution and relative merits of hard X-ray phase-contrast imaging methods}, Phil. Trans. R. Soc. A {\bf 372}, 20130021 (2014). 

E. Wolf, {\em Three dimensional structure determination of semi-transparent objects by holographic data}, Opt. Commun. {\bf 1}, 153--156 (1969). 

E. Wolf, {\em New theory of partial coherence in the space–frequency domain. Part I: spectra and cross spectra of steady-state sources}, J. Opt. Soc. Am {\bf 72}, 343--351 (1982).

H. Yu, R. Lu, S. Han, H. Xie, G. Du, T. Xiao and D. Zhu, {\em Fourier-Transform Ghost Imaging with Hard X Rays}, Phys. Rev. Lett. {\bf 117}, 113901 (2016).  

B. Yu, L. Weber, A. Pacureanu, M. Langer, C.  Olivier, P. Cloetens and F. Peyrin, {\em Phase retrieval in 3D X-ray magnified phase nano CT: Imaging bone tissue at the nanoscale}, 2017 IEEE 14th International Symposium on Biomedical Imaging (ISBI 2017), 56--59 (2017).

M.-C. Zdora, {\em State of the art of X-ray speckle-based phase-contrast and dark-field imaging}, J. Imaging {\bf 4}, 60 (2018).

F. Zernike, {\em Phase contrast, a new method for the microscopic observation of transparent objects}, Physica {\bf 9}, 686--698 (1942). 

A.-X. Zhang, Y.-H. He, L.-A. Wu, L.-M. Chen and B.-B. Wang, {\em Tabletop x-ray ghost imaging with ultra-low radiation}, Optica {\bf 5}, 374--377 (2018).  

P. Zhu, K. Zhang, Z. Wang, Y. Liu, X. Liu, Z. Wu, S. A. McDonald, F. Marone and M. Stampanoni, {\em Low-dose, simple, and fast grating-based X-ray phase-contrast imaging}, Proc. Natl. Acad. Sci. U.S.A. {\bf 107},
13576--13581 (2010).



\end{document}